\documentclass[lettersize,journal]{IEEEtran}
\usepackage{amsmath,amsfonts}
\usepackage{algorithm}
\usepackage{array}
\usepackage[caption=false,font=normalsize,labelfont=sf,textfont=sf]{subfig}
\usepackage{textcomp}
\usepackage{stfloats}
\usepackage{url}
\usepackage{verbatim}
\usepackage{graphicx}
\usepackage{cite}
\usepackage{hyperref}
\usepackage{xcolor} 

\usepackage{anyfontsize}

\usepackage{colortbl}
\usepackage[normalem]{ulem}
\useunder{\uline}{\ul}{}
\usepackage{ulem} 
\usepackage{adjustbox}
\usepackage{xcolor}
\usepackage{multirow}
\usepackage{booktabs} 
\usepackage{algorithmicx}  
\usepackage{algpseudocode}  
\usepackage{amsmath}  

\usepackage{lineno,hyperref}
\begin{document}
\nocite{*}

\title{Towards Structure-aware Model for Multi-modal Knowledge Graph Completion}

\author{Linyu Li,~\IEEEmembership{Student Member,~IEEE,}
\IEEEauthorblockN{Zhi Jin\IEEEauthorrefmark{2},~\IEEEmembership{Fellow,~IEEE,}
Yichi Zhang, Dongming Jin, Chengfeng Dou, Yuanpeng He, Xuan Zhang, Haiyan Zhao}

\thanks{This work was supported by the National Natural Science Foundation of China under Grant No.62436006.

Linyu Li, Zhi Jin, Dongming Jin, Chengfeng Dou,Yuanpeng He and Haiyan Zhao are with Key Laboratory of High Confidence Software Technologies (PKU), Ministry of Education, and School of Computer Science, Peking University, Beijing 100871, China (e-mail: xltx\_youxiang@qq.com; zhijin@pku.edu.cn; dmjin@stu.pku.edu.cn;  chengfengdou@pku.edu.cn ; heyuanpeng@stu.pku.edu.cn; zhhy.sei@pku.edu.cn). Yichi Zhang is with College of Computer Science and Technology, Zhejiang University, Hangzhou 310000, China (e-mail: zhangyichi2022@zju.edu.cn). Xuan Zhang and Jishu Wang are with School of Software, Yunnan Key Laboratory of Software Engineering, Yunnan University, Kunming 650091, China (e-mail: zhxuan@ynu.edu.cn; cswangjishu@hotmail.com).

\textit{© 2025 IEEE. Personal use of this material is permitted. Permission from IEEE must be obtained for all other uses, in any current or future media, including reprinting/republishing this material for advertising or promotional purposes, creating new collective works, for resale or redistribution to servers or lists, or reuse of any copyrighted component of this work in other works.}

}
}



\maketitle
\begin{abstract}
Knowledge graphs (KGs) play a key role in promoting various multimedia and AI applications. However, with the explosive growth of multi-modal information, traditional knowledge graph completion (KGC) models cannot be directly applied. This has attracted a large number of researchers to study multi-modal knowledge graph completion (MMKGC). Since MMKG extends KG to the visual and textual domains, MMKGC faces two main challenges: (1) how to deal with the fine-grained modality information interaction and awareness; (2) how to ensure the dominant role of graph structure in multi-modal knowledge fusion and deal with the noise generated by other modalities during modality fusion. To address these challenges, this paper proposes a novel MMKGC model named TSAM, which integrates fine-grained modality interaction and dominant graph structure to form a high-performance MMKGC framework. Specifically, to solve the challenges, TSAM proposes the Fine-grained Modality Awareness Fusion method (FgMAF), which uses pre-trained language models to better capture fine-grained semantic information interaction of different modalities and employs an attention mechanism to achieve fine-grained modality awareness and fusion. Additionally, TSAM presents the Structure-aware Contrastive Learning method (SaCL), which utilizes two contrastive learning approaches to align other modalities more closely with the structured modality. Extensive experiments show that the proposed TSAM model significantly outperforms existing MMKGC models on widely used multi-modal datasets.

\end{abstract}

\begin{IEEEkeywords}
knowledge graph, knowledge graph completion, multi-modal knowledge graph completion,  Contrastive Learning, link prediction.
\end{IEEEkeywords}

\section{Introduction}\label{chap:1}
\IEEEPARstart{K} NOWLEDGE Graphs (KG) \cite{liang2024survey}\cite{ji2021survey}\cite{ni2023psnea} are a structured form of knowledge representation and currently one of the most popular research areas in the field of knowledge engineering. KGs play a pivotal role in various applications, such as recommendation systems \cite{yi2021multi}\cite{wu2022graph}\cite{cao2022cross}\cite{yang2023knowledge}, social media\cite{cao2022building}, object detection\cite{yang2023context} and applications combined with large language models \cite{pan2024unifying}\cite{zheng2024adapting}. Yet, employing traditional knowledge graphs is no longer adequate to address the escalating and pressing demands of knowledge engineering. The emergence of multi-modal knowledge graphs(MMKGs) \cite{chen2024knowledge}\cite{zhu2022multi}, which additionally links modalities such as images and text, has significantly alleviated this situation. 
\begin{figure}[t!] 
\centering

\includegraphics[width=3.5in]{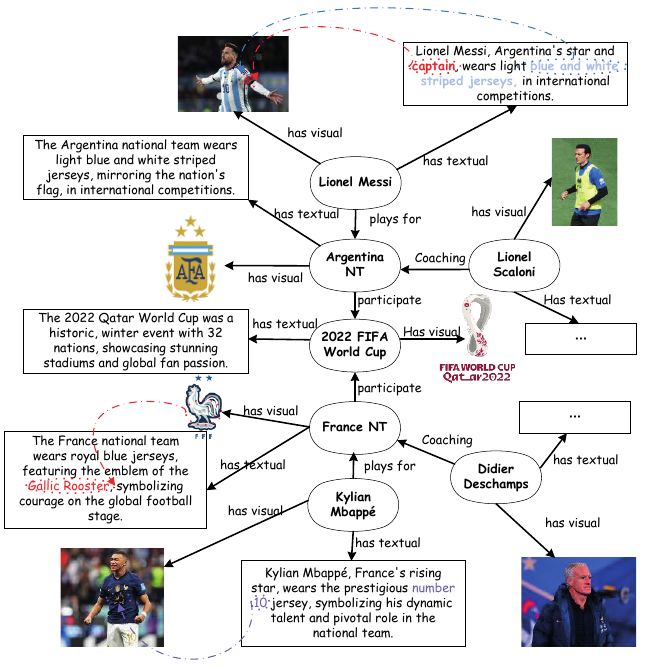}
\caption{A simple example of MMKGC, which includes not only the structural modality but also visual and textual modalities. There are fine-grained interactions between modalities; for instance, the different modalities linked by various dashed lines represent semantically similar meanings at a fine-grained level.}
\label{figure1}
\end{figure}

However, existing MMKGs, like traditional knowledge graphs, suffer from severe incompleteness issues. Multi-modal knowledge Graph Completion (MMKGC) aims to utilize existing multimodal knowledge (text, images, triples, etc.) to obtain a more comprehensive knowledge representation and to predict missing elements in the multimodal knowledge graph to complete it.

Traditional KGC methods \cite{bordes2013translating}\cite{sun2019rotate}\cite{wang2022simkgc}\cite{cao-etal-2021-missing}\cite{shang2023relation} primarily focus on completing static KGs with a single modality and are unable to handle multi-modal KGs. They cannot process the multi-modal attributes (e.g., visual) shown in MMKGC, as illustrated in Fig.1. Therefore, research on MMKGC models is crucial. Recently, significant advancements have been made in MMKGC, with numerous influential studies emerging and achieving certain results. For example, MMKGC models using siamese  networks and multi-hop reason\cite{wei2024multi}
and considering better integration of various modal information:\cite{zhang2022multimodal}\cite{zhang2024native}\cite{zhao2024contrast}\cite{zhang2024mygo}. However, the existing MMKGC work still faces the following two severe challenges.

Firstly, a severe lack of interaction and awareness of fine-grained modality information exists. Typically, existing MMKGC methods embed the multi-modal information of entities into different feature spaces using various embedding models. Subsequently, these multi-modal entity embeddings are fused through operations such as concatenation, averaging, or tokenization to generate a comprehensive embedding representation. This final triple embedding representation is intended to serve as a unified representation of the multi-modal entities. This approach often leads to the model's inability to effectively capture the fine-grained interactions between images and text, as well as a lack of fine-grained modality perception. As illustrated in Fig.1, there are interactions between text and images at a fine-grained level, and different modalities exhibit varying degrees of perceptual fusion. To this end, how to better capture and perceive this fine-grained modal information to assist the structural modality is the focus of the MMKGC task. From a multi-modal perspective, tokenization \cite{radford2021learning}\cite{peng2022beit}\cite{zhang2024mygo} can be understood as a process of transforming data from different modalities into a unified representation of token sequences. This unified representation helps eliminate feature disparities across modalities, enabling multi-modal models to process and integrate data within a shared feature space. Consequently, it facilitates the comprehensive exploration of synergistic information and fine-grained interactions between modalities. This situation highlights the urgent need to explore how tokenization can be utilized to capture fine-grained interactions within each modality and enhance awareness of modality-specific knowledge, thereby enabling more effective integration of multi-modal feature representations.

\begin{figure}[!t]
\centering
\includegraphics[width=3.5in]{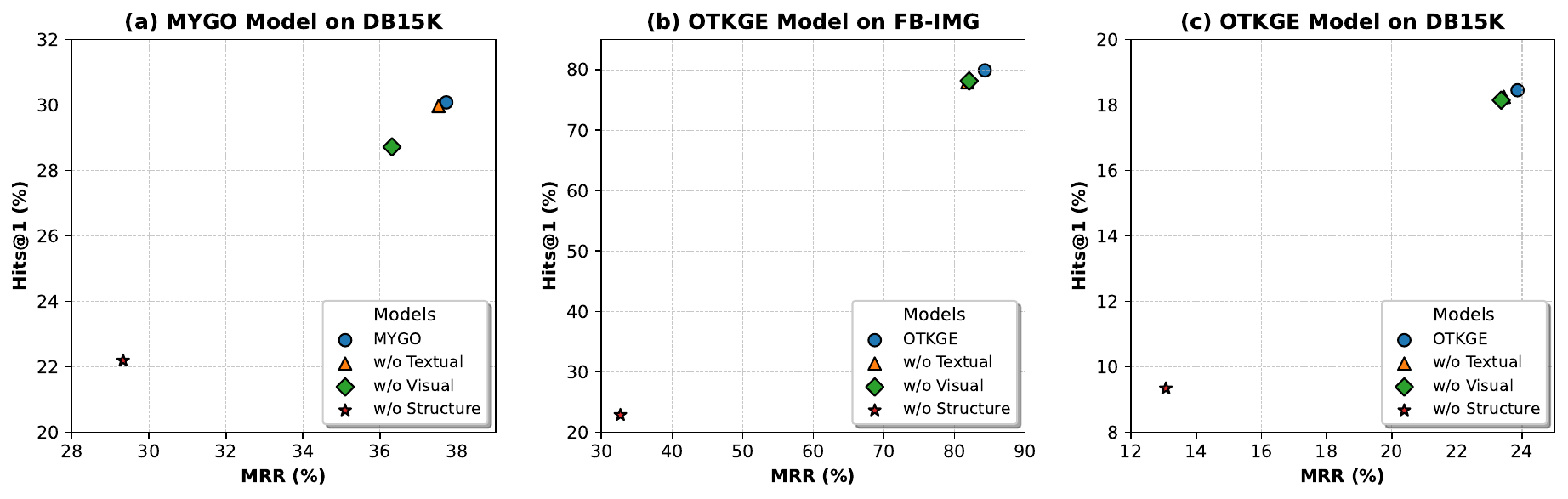}
\caption{Performance of the MyGO\cite{zhang2024mygo} and OTKGE\cite{cao2022otkge} models in terms of MRR and Hits@1 after completely removing all modality knowledge.}

\label{mike_amp}
\end{figure}

Secondly, we discovered a significant issue: existing MMKGC models severely underestimate the dominant role of graph structure in the fusion of multi-modal knowledge. As shown in Fig.2, we conducted verification tests on two typical open-source MMKGC models, MyGO\cite{zhang2024mygo} and OTKGE\cite{cao2022otkge}, after completely removing the structural modal knowledge. Without the support of structural modality knowledge, metrics such as MRR and Hits@1 exhibit a dramatic decline. 
Theoretically, structural modality explicitly captures the intrinsic relationships between entities, forming the backbone of knowledge graph reasoning. It provides a structured and semantically rich representation for entity interactions, which is crucial for reasoning and task completion. In contrast, textual and visual modalities merely serve as auxiliary components to enhance the structured knowledge for the reasoning process.
Moreover, the fusion of multi-modal knowledge is often accompanied by noise \cite{ngiam2011multimodal}\cite{gao2024embracing}. Since embeddings from different modalities typically exist in distinct heterogeneous spaces, even when using mapping operations like those in \cite{zhu2023minigpt} for fusion, the original distribution characteristics of each modality's embeddings can still be disrupted, leading to inconsistencies in representations within a unified space. Similarly, 
due to the above situation, it is urgent for us to study an alignment method that aligns other modalities with the structural modality and fully explore the dominant role of the structural modality in MMKGC, so as to reduce the noise generated in the process of multi-modal knowledge fusion.

To address the aforementioned challenges, this paper proposes a novel MMKGC model named TSAM, which is designed towards structure-aware modeling. TSAM comprises two core methods: Fine-grained Modality Awareness Fusion (FgMAF) and Structure-aware Contrastive Learning (SaCL). The FgMAF method first utilizes visual pre-trained models \cite{bao2021beit}\cite{peng2022beit} and text pre-trained models \cite{devlin2018bert}\cite{liu2019roberta}\cite{he2020deberta} to perform tokenization on the visual and textual modalities of entities in the MMKG, capturing fine-grained semantic token sequences for each modality. Then, using a transformer-based\cite{vaswani2017attention} approach, it encodes the obtained entity sequences from different modalities. This is followed by a modality attention mechanism combined with a decoding operation to perceptively and interactively capture the multi-modal information within the MMKG. The SaCL method, on the other hand, incorporates two joint contrastive learning paradigms. Through contrastive learning, TSAM learns to align visual and textual representations with structured representations, reducing noise in the vector representations of other modalities after fusion, thereby bringing them closer to the vector space and enhancing the effectiveness of MMKGC. Additionally, TSAM employs KGE models such as \cite{bordes2013translating}\cite{sun2019rotate}\cite{balavzevic2019tucker} to serve as scoring functions and capture structural-semantic relationships, obtaining structured embeddings. The core contributions of this paper can be summarized as follows:
\begin{itemize}
    \item  This paper introduces the Fine-grained Modality Awareness Fusion (FgMAF) method, which captures interactions between different modalities at the finest granularity level and uses a modality attention mechanism to perceive semantic information across various modalities.
    \item To the best of our knowledge, this is the first work that systematically analyzes and emphasizes the critical importance of structural modality in the field of multi-modal knowledge graph completion. Furthermore, we propose the Structure-aware Contrastive Learning (SaCL) method, which effectively aligns other modalities with the structural modality. This approach mitigates the noise introduced to the structural modality during modality fusion, under the premise that the structural modality remains dominant.
    \item Through comprehensive experiments on three real-world benchmark datasets, we have thoroughly demonstrated the effectiveness of our model. Compared to other MMKGC models, TSAM achieved optimal performance across all metrics on both datasets.
\end{itemize}

The remainder of this paper is organized as follows: In Section II, we review related work on Knowledge Graph Completion (KGC). Section III provides a formal definition of the research problem and presents a detailed introduction to the TSAM model. Section IV and Section V present the experimental setup and the experimental results. Finally, in Section VI, this paper summarizes the proposed TSAM model.


\section{Related Work}\label{chap:2}
\subsection{Non-multi-modal KGC model}
In general, non-multi-modal KGC models include the following types. 

\textbf{Traditional embedding models based on scoring functions}: By designing scoring functions in various vector spaces to constrain the distance between head and tail entities to optimize model representation, such as TransE \cite{bordes2013translating}, TransR \cite{lin2015learning}, RotatE\cite{sun2019rotate}, HAKE \cite{zhang2020learning}, QIQE \cite{li2023knowledge}, WeightE \cite{zhang2023weighted}, ConKGC \cite{shang2023relation}, GIE \cite{cao2022geometry}, SpherE \cite{li2024sphere}, MRME \cite{li2025multi}, RecPiece \cite{liangclustering}, ExpressivE \cite{pavlovicexpressive} and other models. These embedding-based models constrain the distance between the head entity and the tail entity by designing scoring functions in different vector spaces, thereby continuously optimizing the representation of entities and relations in KG to capture the latent semantic relationships between entities and relations in KG and achieve the purpose of KGC.

\textbf{ Models based on natural language processing}: By converting triples into text sequences, using transformer-based models to perform encoder-decoder operations to achieve prediction, such as SimKGC\cite{wang2022simkgc}, KG-Bert\cite{yao2019kg}, CSProm-KG \cite{chen2023dipping}, StAR \cite{wang2021structure}, and other models. The commonality of this type of model is that it fully combines the semantic understanding ability of natural language processing with the structured characteristics of knowledge graphs by converting structured knowledge graphs into continuous text sequences. Its core advantages are: first, using pre-trained language models (such as BERT) to capture deep semantic associations; second, through flexible sequence generation or contrastive learning strategies, it enhances the ability to reason about complex relationships.

\textbf{Models based on graph neural networks(GNN):} By using KG completely in the form of GNN as an encoder to perform link prediction tasks to achieve the purpose of KGC, such as CompGCN \cite{vashishth2019composition}, CLGAT\cite{li2022knowledge}, NBFNet \cite{zhu2021neural}, InGram \cite{pmlr-v202-lee23c}, MGTCA \cite{shang2024mixed}, and other models. The GNN-based model takes the topological structure and neighborhood relationship of the knowledge graph as the core, and explicitly models the multi-hop semantic associations between entities. Although this type of method has significant advantages in modeling complex relationship paths, the computational efficiency and long path dependency issues are still areas that need to be improved.

\subsection{Multi-modal KGC model}

MMKGC \cite{chen2024knowledge}\cite{zhu2022multi}\cite{liang2024survey2} enhances missing entity prediction by leveraging auxiliary modalities like text and images to complement structural information. Existing methods tackle key challenges such as modality alignment, imbalanced multi-modal fusion, and noisy or missing modality data through diverse strategies. Typical MMKGC methods, such as:  OTKGE \cite{cao2022otkge}, MyGo\cite{zhang2024mygo},  LAFA \cite{shang2024lafa}, MR-MKG \cite{lee2024multimodal}, IMF \cite{li2023imf}, SGMPT \cite{liang2024simple}, CMR \cite{zhao2024contrast}, SGMPT\cite{liang2024simple}, DySarl \cite{liu2024dysarl}, MKGformer \cite{chen2022hybrid} and MGKsite\cite{liang2024mgksite}, extend single-modality KGE approaches by integrating multi-modal embeddings, which are extracted via pre-trained models, to optimize predictions and represent entities from diverse perspectives.

For instance,  Alignment and Optimal Transport: OTKGE \cite{cao2022otkge} optimizes cross-modal consistency by minimizing Wasserstein distances between structural and multi-modal embeddings, while CMR \cite{zhao2024contrast} employs contrastive learning to align modality-specific features in a shared latent space. These methods mitigate heterogeneity across modalities, enhancing graph completion robustness.
Dynamic Fusion and Attention: To address imbalanced modality contributions, LAFA \cite{shang2024lafa} introduces attention mechanisms that adaptively weight modalities based on relational contexts. MyGO \cite{zhang2024mygo} extends this with cross-modal entity encoding and fine-grained contrastive learning, capturing nuanced entity relationships. NativE \cite{liu2024dysarl} further innovates with relation-guided dual adaptive fusion, prioritizing modalities most relevant to specific triples. Adversarial and Contrastive Learning: AdaMF \cite{shang2024lafa} integrates adversarial training to balance underrepresented modalities, demonstrating robustness against data imbalances. SGMPT \cite{liang2024simple} and MMRNS \cite{li2023imf} leverage contrastive learning with semantic-aware negative sampling, refining discriminative power in entity disambiguation. Transformers and Cross-modal Integration: Transformer-based architectures like VISTA \cite{lee2023vista} and MKGformer \cite{chen2022hybrid} excel in joint image-text representation learning, decoding complex cross-modal interactions for state-of-the-art performance. SnAg \cite{liu2024dysarl} enhances noise robustness through modality-level masking, ensuring reliable fusion even with incomplete data. Multi-stage Fusion Frameworks: IMF \cite{li2023imf} adopts a two-stage approach, preserving modality-specific knowledge via bi-linear pooling before integrating complementary embeddings, effectively balancing specificity and generality.


\section{METHODOLOGY}
In this section, we first present the formal definitions related to MMKGC, followed by a detailed explanation of the TSAM model's intricacies. This includes the two core methods: FgMAF and SaCL. Finally, we will describe the detailed process of model training and the loss function.

\subsection{Preliminary and Task Formulation}
Formally speaking, MMKG can be defined as: $\mathcal{G}=(\mathcal{E}, \mathcal{R}, \mathcal{T}, \mathcal{M})$ where $\mathcal{E}$ and $\mathcal{R}$ are entity sets and relation sets respectively. $\mathcal{T}=\{(h, r, t) \mid h, t \in \mathcal{E}, r \in \mathcal{R}\}$ represents a triplet where a head entity $h$ is connected to a tail entity $t$ through a relation $r$. In addition, $\mathcal{M}=\{\mathbf{S} \cup \mathbf{V} \cup \mathbf{T} \}$ corresponds to the structural modal information $\mathbf{S}$, visual modal information $\mathbf{V}$ and textual modal information $\mathbf{T}$ of each entity $e$.

The structural modality $\mathbf{S}$ refers to the intrinsic graph structure of the knowledge graph. It is captured by the set of triples $(h, r, t)$ and encodes the relational and connectivity information among entities. This modality is typically learned using knowledge graph embedding techniques (e.g., TransE \cite{bordes2013translating}, TuckER \cite{balavzevic2019tucker},RotatE \cite{sun2019rotate}) and serves as the backbone of our entity representations.
The visual modality $\mathbf{V}$ corresponds to the visual data (such as images) associated with entities. Using pre-trained visual encoders (e.g., BEIT-V2\cite{bao2021beit}), the images are tokenized into fine-grained visual tokens, which capture the semantic content and nuances of the visual information.
The textual modality $\mathbf{T}$ consists of textual descriptions, captions, or other text associated with entities. Pre-trained language models (e.g., BERT \cite{devlin2018bert}) are employed to tokenize and encode this textual data into discrete tokens, thereby extracting the underlying semantic features.

The core task of KGC can be formalized as a connection prediction task, for example: given a missing triple $(h,r,?)$, predict the missing tail entity $t$ by giving the head entity $h$ and relation $r$. And construct a score function $\mathbf{Score}(h, r, t): \mathcal{E} \times \mathcal{R} \times \mathcal{E} \rightarrow \mathbb{R}$ to quantitatively score the rationality of the triple $(h, r, t)$. Slightly different from KGC, MMKGC further considers the multi-modal information $\mathcal{M}$ of each entity in the entity set $\mathcal{E}$ to enhance the embedding representation of MMKG and achieve better results.

\begin{figure*}[ht!] 
\centering
\includegraphics[width=7in]{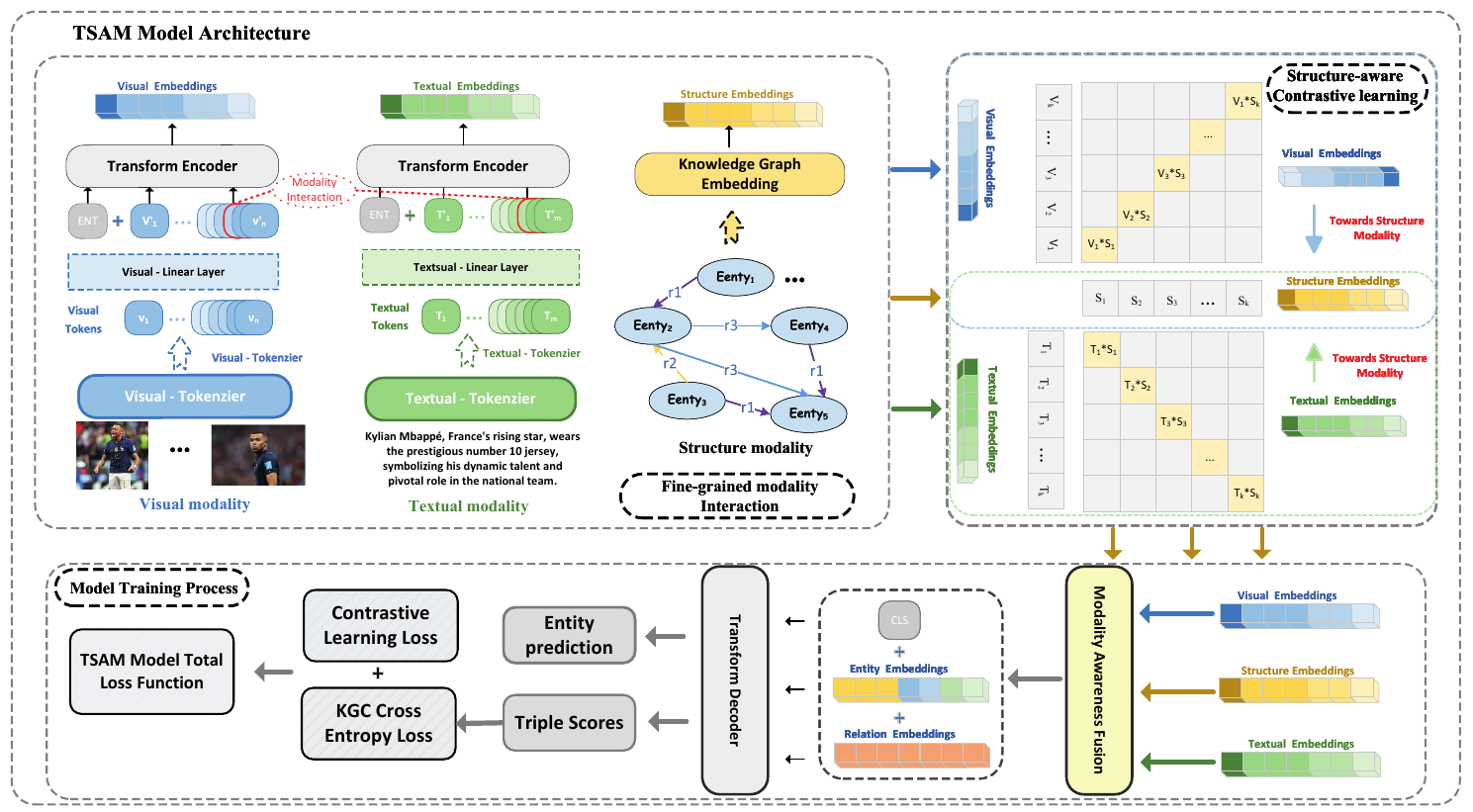}

\caption{The Architecture of the TSAM Model. TSAM incorporates the FgMAF method to better fuse and perceive various modalities
knowledge in MMKG, while the SaCL method is employed to align other modal knowledge with the structural modality, with the
structural modality as the dominant factor. TSAM employs iterative entity representation updates and contrastive learning to achieve representation learning. The optimized entity and relation representations are then input into the scoring function to perform relevant triplet link prediction.}
\label{figure3}
\end{figure*}


\subsection{Architecture Overview}
The overall architecture of the TSAM model is illustrated in Fig.3, which is a novel MMKGC model designed to enhance multi-modal performance in MMKGC by effectively integrating fine-grained modality interaction awareness and emphasizing the dominant role of the structural modality. The model consists of two primary components: the FgMAF method and the SaCL method. FgMAF uses pre-trained models to tokenize visual and textual data into discrete tokens, which are then linearly projected into a unified space and encoded via a Transformer-based encoder. An attention mechanism is applied to weigh the importance of each modality, ensuring focus on the most relevant information during fusion. SaCL incorporates contrastive learning to align visual and textual embeddings with the structural modality, reducing noise and enhancing structural integrity. A Transformer-based decoder predicts the tail entity in the triple, with a combination of cross-entropy loss for prediction and contrastive learning loss for modality alignment. The total loss function is optimized during training to improve model performance.


\subsection{Fine-grained modality awareness Fusion}
Consistent with previous mainstream MMKGC works \cite{cao2022otkge}\cite{zhang2024native}\cite{zhang2024mygo}\cite{li2023imf}, this paper also considers three types of modality information in MMKGC: visual, textual, and structural modalities.

\textbf{Visual Tokenizer}{:} In order to capture fine-grained interaction information at the token level, this paper follows the setting in previous work \cite{zhang2024mygo}. We use the visual pre-trained model BEIT-V2 \cite{peng2022beit} to convert the image corresponding to each entity $e$ into a set of discrete visual tags through a visual tagger, each of which corresponds to an image patch. Thus, the visual tag $\mathbf{Tokens}_{visual}(e)$ of each entity $e$ is obtained:
\begin{equation}
        \mathbf{Tokens}_{visual}(e)=\text{V-Encoder(e)}= \left\{v_{1}, v_{2}, \cdots, v_{n}\right\}
\end{equation}
where $n$ represents the number of visual modality tokens. $\text{V-Encoder(e)}$ represents the visual encoder \cite{peng2022beit}.

\textbf{Textual Tokenizer}: Similar to the process of Visual Tokenizer, we use the pre-trained language model Bert \cite{devlin2018bert} to convert the text description paragraph corresponding to each entity $e$ into a set of discrete text tags through the text tagger, where each text tag corresponds to the smallest unit of a text. Thus, we obtain the text tag $\mathbf{T}_{textual}(e)$ of each entity $e$:
\begin{equation}
        \mathbf{Tokens}_{textual}(e)=\text{T-Encoder(e)}= \left\{t_{1}, t_{2}, \cdots, t_{m}\right\}
\end{equation}
Where $m$ represents the number of text modality tokens and $\text{T}$-Encoder(e) represents the text encoder \cite{devlin2018bert}.

\textbf{Visual and Textual Encoder}: After obtaining the fine-grained information tokens of visual and textual, TSAM is different from MyGo \cite{zhang2024mygo} which directly concatenates the tokens of the two modalities. TSAM aims to directly interact and perceive the fine-grained information between modalities from a more fine-grained perspective. First, we define two linear projection layers $g_{v}\left(\right)$ and $g_{t}\left(\right)$ to project the visual and textual tokens into the same space:

\begin{small}
    \begin{equation}
    \mathbf{Tokens}^{'}_{visual}(e)=\{{v}^{'}_{1},...,{v}^{'}_{n}\}=\{g_{v}\left(v_{1}\right)+b^{v}_{1} ,...,g_{v}\left(v_{n}\right)+b^{v}_{n} \}
    \end{equation}
\end{small}
\begin{small}
    \begin{equation}
    \mathbf{Tokens}^{'}_{txtual}(e)=\{{t}^{'}_{1},...,{t}^{'}_{n}\}=\{g_{t}\left(t_{1}\right)+b^{t}_{1} ,...,g_{t}\left(t_{n}\right)+b^{t}_{n} \}
    \end{equation}
\end{small}%
where $\{{t}^{'}_{1},...,{t}^{'}_{n}\}$ represent the tokens that have been transformed into the unified spatial dimension through linear projection, $b^v$ and $b^t$ represent bias vectors, so as to better integrate the two modalities through training the linear projection layer. After obtaining the token sequences of the two modalities after linear layer projection, TSAM uses the pre-trained language model based on transformer\cite{vaswani2017attention} to perform encoder processing on the sequences respectively:

\begin{equation}
    \mathrm{e}_{vis}=\operatorname{Pooling}\left(g_{e}\left([\mathrm{ENT}],  v_{1}^{'} , \cdots, v_{n}^{'}\right)\right)
\end{equation}
\begin{equation}
    \mathrm{e}_{txt}=\operatorname{Pooling}\left(g_{e}\left([\mathrm{ENT}],  t_{1}^{'} , \cdots, t_{n}^{'}\right)\right)
\end{equation}
where $g_{e}$ represents the encoder layer based on the Transformer \cite{vaswani2017attention} pre-trained language model, Pooling is the pooling operation, and [ENT] is similar to [CLS] in Bert\cite{devlin2018bert}, which is used to obtain the final hidden representation of the token. $\mathrm{e}_{vis}$ $\mathrm{e}_{txt}$ represent the visual and text embeddings of entity $e$ respectively.

\textbf{Structural Encoder: } Typically, the semantic information of structural modality is learned through triples through the knowledge graph embedding (KGE) model. These embeddings are learned during training by optimizing scoring functions that capture the semantic relationships in the knowledge graph. KGE uses a scoring function to evaluate the authenticity of the triple $(h, r, t)$. 
In KGE models (e.g., TransE \cite{bordes2013translating}, TuckER \cite{balavzevic2019tucker}, RotatE \cite{sun2019rotate}), the embeddings for entities $\mathbf{h}, \mathbf{t} \in \mathbb{R}^d$ and relations $\mathbf{r} \in \mathbb{R}^d$ are initialized as trainable vectors. These embeddings form the basis for the structural modality.
During training, these embeddings are optimized by minimizing a loss function (e.g., margin-based loss) that ensures plausible triples $(h, r, t)$ receive higher scores than implausible ones.
The scoring function is not only used to evaluate the interaction between entities and relations but also plays an important role in learning structural information. We adopt the TuckER \cite{balavzevic2019tucker}, TransE \cite{bordes2013translating} and RotatE \cite{sun2019rotate} models to construct a structured encoder to achieve more accurate knowledge representation.

For TuckER \cite{balavzevic2019tucker}, its core idea is to use Tucker decomposition to represent the score of the triple as a product of a tensor. Specifically, TuckER defines the scoring function of the triple $(h, r, t)$ in the knowledge graph as:
\begin{equation}
\operatorname{Score}(h, r, t)=\sum_{i, j, k} \mathbf{W}_{i j k} \cdot \mathbf{h}_i \cdot \mathbf{r}_j \cdot \mathbf{t}_k
\end{equation}
where $\mathbf{W}_{ijk}$ is a learnable three-dimensional tensor weight, which represents the interaction weight of the relation $r$ on the head entity $h$ and the tail entity $t$, $\mathbf{h}_i$, $\mathbf{r}_j$ and $\mathbf{t}_k$ are the embedding vectors of the head entity, relation and tail entity respectively. 

For TransE\cite{bordes2013translating}, the core idea is to capture the relation between entities through vector addition. The model assumes that each relation can be regarded as a "translation" operation, that is, given the head entity $h$ and the relation $r$, the tail entity $t$ can be obtained by "translating" the head entity in the direction of the relation. Specifically, the basic formula of TransE is:
\begin{equation}
\operatorname{Score}(h, r, t)=\|\mathbf{h}+\mathbf{r}-\mathbf{t}\|
\end{equation}

For RotatE \cite{sun2019rotate}, its core idea is to model relations as rotations in a complex vector space. The method represents the relationship between entities by rotating the head entity vector in the complex plane. Specifically, RotatE defines the scoring function of the triple $(h, r, t)$ as:
\begin{equation}
\operatorname{Score}(h, r, t) = |\mathbf{h} \circ \mathbf{r} - \mathbf{t}|
\end{equation}
where $\mathbf{h}, \mathbf{r}, \mathbf{t} \in \mathbb{C}^d$ are complex-valued embeddings of the head entity, relation, and tail entity, respectively. The operator $\circ$ denotes the Hadamard (element-wise) product, which performs a rotation operation on $\mathbf{h}$ in complex space. The score measures the distance between the rotated head entity $\mathbf{h} \circ \mathbf{r}$ and the tail entity $\mathbf{t}$, enforcing geometric consistency in the embedding space.

These scoring functions guide the learning of structural embeddings through backpropagation, ensuring that semantically valid triples are assigned higher scores. Through the above three KGE models, we have the embedding representation of the structural modality of entity $e$: $e_{str}$ and the embedding representation of the relation $r$.

\textbf{Modality Awareness Fusion}: In our model, we introduce a Modality Awareness Fusion mechanism to effectively integrate the embedding representations of an entity 
$e$ across three distinct modalities: structural, visual, and textual. This fusion process is designed to leverage the complementary information provided by each modality, thereby enhancing the overall representation of the entity. After obtaining the embedding representations $e_{str}$, $\mathrm{e}_{vis}$ and $\mathrm{e}_{txt}$, and for the entity $e$ in the structural, visual, and textual modalities, respectively, we construct an attention vector to dynamically weigh the importance of each modality in the final fused representation. The attention mechanism is crucial as it allows the model to focus on the most informative aspects of each modality, thereby improving the quality of the fused representation.The fused entity representation $e_{f}$ is computed as follows:

After obtaining the embedding representations of entity $e$ in three modalities, we construct an attention vector to enhance modality fusion: 

\begin{equation}
e_{f}=\operatorname{stack}\left(\alpha_s \boldsymbol{e}_{str}, \alpha_v \boldsymbol{e}_{vis}, \alpha_t \boldsymbol{e}_{txt} \right)
\end{equation}
\begin{equation}
    \left(\alpha_s, \alpha_v, \alpha_t\right)=\operatorname{Softmax}\left(\boldsymbol{\alpha}^T e_{vis}, \boldsymbol{\alpha}^T e_{txt}, \boldsymbol{\alpha}^T \boldsymbol{e}_{str}\right)
\end{equation}    
where $\boldsymbol{\alpha}$ is the attention vector and  $e_{f}$ represents the fused entity representation. The attention weights 
$\alpha_s, \alpha_v, \alpha_t$ reflect the relative importance of the structural, visual, and textual modalities, respectively, in the context of the entity $e$. These weights are normalized using the softmax function to ensure that they sum to one, providing a probabilistic interpretation of the modality contributions. This fine-grained interaction allows the model to perceive subtle inter-modal relationships, which are crucial for accurately representing the entity in a multi-modal context.

Finally, we learn the entity representation $e_{f}$ with very fine-grained interaction and inter-modal perception. through this modality-aware fusion process, our model is able to achieve a high level of granularity in capturing the nuances of each modality while also integrating them in a coherent and meaningful way. This approach ensures that the final representation is both rich and contextually relevant, enabling superior performance in downstream tasks that require multi-modal understanding.

\subsection{Structure-aware Contrastive learning}
Although modality fusion can be achieved through linear transformations and attention mechanisms, a semantic gap invariably exists between different modalities. Contrastive learning \cite{chen2020simple}\cite{he2020mocov1}\cite{radford2021learning}\cite{gao2021simcse}\cite{liang2023knowledge} has garnered significant attention across various fields, as it enhances the representation of similar samples by bringing them closer together while pushing dissimilar samples apart. This paper aims to align the visual and textual modality representations with the structural modality through contrastive learning, thereby mitigating noise potentially introduced by irrelevant images and texts, and ultimately improving the model's predictive performance. Specifically, to achieve this effect, the SaCL method performs contrastive learning twice, centering on the structured modality.

In the contrastive learning of structural modality-visual modality, the embedding set $ S = \{s_{1}, \dots, s_{k}\}$ of the entity structural modality $E_{str}$ and the embedding set $ V = \{v_{1}, \dots, v_{k}\}$ of the entity visual modality $E_{vis}$ are positive samples of each other. And randomly select K other samples in the same mini-batch as negative sample pairs of $E_{str}$ and $E_{vis}$, and express them as:
\begin{equation}
    V_{\boldsymbol{i}}^{-}=\left\{V_{i 1}^{-}, V_{i 2}^{-}, \ldots, V_{i K}^{-}\right\} , S_{\boldsymbol{i}}^{-}=\left\{S_{i 1}^{-}, S_{i 2}^{-}, \ldots, S_{i K}^{-}\right\}
\end{equation}

After that, the negative log-likelihood function is used to train the maximum similarity between positive sample pairs and the minimum similarity between negative sample pairs. The formula is as follows:
\begin{small}
    \begin{equation}
        \mathcal{L}_{S_V}=-\frac{1}{B} \sum_{i=1}^B \log \frac{\exp \left(s\left(S_i, V_i\right) / \tau\right)}{\exp \left(s\left(S_i, V_i\right) / \tau\right)+\sum_{j=1}^K \exp \left(s\left(S_i, V_{i j}^{-}\right) / \tau\right)}
    \end{equation}
    \begin{equation}
        \mathcal{L}_{V_S}=-\frac{1}{B} \sum_{i=1}^B \log \frac{\exp \left(s\left(V_i, S_i\right) / \tau\right)}{\exp \left(s\left(V_i, S_i\right) / \tau\right)+\sum_{j=1}^K \exp \left(s\left(V_i, S_{i j}^{-}\right) / \tau\right)}
    \end{equation}
\end{small}
\begin{equation}
        \mathcal{L}_{SV}=\mathcal{L}_{S_V} +\mathcal{L}_{V_S}
\end{equation}
where $B$ represents the total batch size, $s(\cdot)$ represents the calculation of the cosine similarity of two tensors, and $\tau $ is the temperature parameter. $\mathcal{L}_{SV}$ represents the contrast loss of the structure-visual modality so that the visual modality and the textual modality can be better aligned and reflect the more realistic graph structure pattern in MMKG.

In the contrastive learning of structural modality and text modality, the embedding set $ T = \{t_{1}, \dots, t_{k}\}$ of entity text modality $E_{txt}$ and the embedding set $ S = \{s_{1}, \dots, s_{k}\}$ of $E_{str}$ are positive samples of each other. And K other samples in the same mini-batch are randomly selected as negative sample pairs of $E_{str}$ and $E_{txt}$ and are expressed as:
\begin{equation}
     S_{\boldsymbol{i}}^{-}=\left\{S_{i 1}^{-}, S_{i 2}^{-}, \ldots, S_{i K}^{-}\right\} , T_{\boldsymbol{i}}^{-}=\left\{t_{i 1}^{-}, t_{i 2}^{-}, \ldots, t_{i K}^{-}\right\} 
\end{equation}

Similar to the structural modality-visual modality operation, the contrast loss of structural modality-textual modality can be defined as:
\begin{small}
    \begin{equation}
        \mathcal{L}_{S_T}=-\frac{1}{B} \sum_{i=1}^B \log \frac{\exp \left(s\left(S_i, T_i\right) / \tau\right)}{\exp \left(s\left(S_i, T_i\right) / \tau\right)+\sum_{j=1}^K \exp \left(s\left(S_i, T_{i j}^{-}\right) / \tau\right)}
\end{equation}
\begin{equation}
        \mathcal{L}_{T_S}=-\frac{1}{B} \sum_{i=1}^B \log \frac{\exp \left(s\left(T_i, S_i\right) / \tau\right)}{\exp \left(s\left(T_i, S_i\right) / \tau\right)+\sum_{j=1}^K \exp \left(s\left(T_i, S_{i j}^{-}\right) / \tau\right)}
\end{equation}
\end{small}
\begin{equation}
    \mathcal{L}_{ST}=\mathcal{L}_{S_T} +\mathcal{L}_{T_S}
\end{equation}
where $\mathcal{L}_{ST}$ represents the contrastive loss of structural modality-text modality so that the text modality and structural modality can be better aligned and reflect a more realistic graph structure pattern in MMKG.
\subsection{Model Training Process}
After having the fused representation of the entity $\boldsymbol{e_f}$ and the relation embedding representation $r$ obtained using KGE, we use the Transformer-based decoder to obtain the prediction result of the tail entity in the triple:
\begin{equation}
    \boldsymbol{t}^{p}=g_{d}([CLS] , \boldsymbol{h_f} , \mathbf{r})
\end{equation}
where $g_{d}()$ represents the decoder layer with a Transformer-based pre-trained language model \cite{devlin2018bert}\cite{liu2019roberta}\cite{he2020deberta}, $\boldsymbol{h_f}$ represents the modality fusion representation of the head entity. $[CLS]$ indicates that the final representation of the token is used as the overall representation of the input sequence and is predicted and classified through a fully connected layer. $\boldsymbol{t}^{p}$ represents the predicted tail entity given $\boldsymbol{h_f}$ and $\boldsymbol{r}$.

We choose the cross entropy loss function as the core loss function for model prediction, which is defined as follows:
\begin{equation}
\begin{split}
\mathcal{L}_{p} = \sum_{(h, r, t) \in \mathcal{T}} -\frac{1}{|\mathcal{E}|} \sum_{n=1}^{|\mathcal{E}|} \left( y \cdot \log \left(\Theta\left(h, r, t_n\right)\right) + \right. \\
\left. \left(1-y\right) \cdot \log \left(1-\Theta\left(h, r, t_n\right)\right) \right)
\end{split}
\end{equation}
\begin{equation}
    \Theta (h, r, t)=\operatorname{sigmoid}(Score(h, r, t))
\end{equation}
Where $\mathcal{E}$ is the entire set of candidate prediction entities, $y \in\{0,1\}$ is the label of the triple $\left(h, r, t_n\right)$. The total loss of the model is:
\begin{equation}
    \mathcal{L}= \mathcal{L}_{p} + \mathcal{L}_{ST} + \mathcal{L}_{SV}
\end{equation}
where $\mathcal{L}_{p}$ represents the cross entropy loss function for prediction. $\mathcal{L}_{ST}$ and $\mathcal{L}_{SV}$ represent the contrastive learning loss functions for modality alignment, respectively.
The training process of the model is shown in Algorithm 1.
\begin{algorithm}
\caption{TSAM Model for multi-modal Knowledge Graph Completion }
\begin{flushleft}
\begin{algorithmic}[1]
\renewcommand{\algorithmicrequire}{\textbf{Input:}}
\renewcommand{\algorithmicensure}{\textbf{Output:}}
\Require Knowledge Graph $\mathcal{G} = (\mathcal{E}, \mathcal{R}, \mathcal{T}, \mathcal{M})$ with multi-modal data $\mathbf{V}$, $\mathbf{T}$, $\mathbf{S}$
\Ensure Optimized embeddings for each modality

\State \textbf{Initialization:} Pre-trained tokenizers, scoring function, attention parameters

\State \textbf{Fine-grained Modality-aware Fusion (FgMAF):}
\For{each entity $e \in \mathcal{E}$}
    \State $\text{Tokens}_{\text{visual}} = \text{V-Encoder}(e)$
    \State $\text{Tokens}_{\text{textual}} = \text{T-Encoder}(e)$
    \State Project tokens $\text{Tokens}_{\text{visual}}$ and $\text{Tokens}_{\text{textual}}$ to unified space 
    \State Use Transformer layers and KGE model to represent multi-modal $\boldsymbol{e}_{vis}$, $\boldsymbol{e}_{txt}$ $\boldsymbol{e}_{str}$ respectively
\EndFor

\State {Modality Fusion:}
\For{each entity $e \in \mathcal{E}$}
    \State Compute $e_{f}$ by combining $\boldsymbol{e}_{str}$, $\boldsymbol{e}_{vis}$, $\boldsymbol{e}_{txt}$, and $\text{Encoded}_{\text{text}}$ with attention mechanism
\EndFor

\State \textbf{Structure-aware Contrastive Learning (SaCL):}
\State Generate positive and negative samples for contrastive learning
\State Compute contrastive losses $\mathcal{L}_{SV}$ for (structural, visual) and $\mathcal{L}_{ST}$ for (structural, textual) pairs

\State Total contrastive loss = $\mathcal{L}_{ST}$ + $\mathcal{L}_{SV}$

\State \textbf{Training:}
\For{each $(h, r, t) \in \mathcal{T}$}
    \State Predict the tail entity $t_{\text{pred}} = \text{Decoder}(h_{\text{fused}}, r)$ and calculate the score function and cross-entropy loss $\mathcal{L}_{ST}$
\EndFor
\State $\text{Total Loss} = \text{prediction loss} + \text{Total}_{\text{contrastive loss}}$
\State \textbf{Repeat:}Training and optimized is repeated to get the best-predicted value
\State \textbf{Until:}Converges
\end{algorithmic}
 
\end{flushleft}
\end{algorithm}

\section{Experiment}
\subsection{Dataset and Evaluation metrics }
\subsubsection{Dataset}
This study employs three of the most widely used and publicly available benchmarks for MMKGC, DB15K \cite{liu2019mmkg} and MKG-W/Y \cite{xu2022relation}, to comprehensively evaluate the model’s performance in multi-modal information fusion. Both datasets consist of three types of modality information: structural triples, entity images, and entity descriptions. Table I provides detailed statistics on the two datasets. 

\begin{table}[h]
\centering
\caption{Statistical of the DB15K \cite{liu2019mmkg} and MKG-W/Y \cite{xu2022relation} Datasets.}
\resizebox{\columnwidth}{!}{%
\begin{tabular}{lccccccc}
\toprule
\textbf{Dataset} & \(|\mathcal{E}|\) & \(|\mathcal{R}|\) & \#Vis & \#Text & \#Train & \#Valid & \#Test \\
\midrule
DB15K & 12,842 & 279 & 12,818 & 12,842 & 79,222 & 9,902 & 9,904 \\
MKG-W & 15,000 & 169 & 14,463 & 14,123 & 34,196 & 4,276 & 4,274 \\
MKG-Y & 15,000 & 28 &  14,244 & 14,305 & 21,310 & 2,665 & 2,663 \\
\bottomrule
\end{tabular}%
}
\end{table}

\begin{table*}[t]
\centering
\caption{The experimental results of TSAM and the baseline model on three MMKG datasets. $\clubsuit$ represents the experimental results that we reproduced through its source code. The rest of the baseline results are from the source papers of their respective models and the report in \cite{zhang2024mygo} and \cite{chen2024power}.}
\begin{tabular}{cccccccccccccc}
\toprule
\multicolumn{2}{c}{}                                                      & \multicolumn{4}{c}{DB15K} & \multicolumn{4}{c}{MKG-W} & \multicolumn{4}{c}{MKG-Y} \\
\cmidrule(lr){3-6} \cmidrule(lr){7-10} \cmidrule(lr){11-14}
\multicolumn{2}{c}{Model}                               & MRR & Hit@1 & Hit@3 & Hit@10 & MRR & Hit@1 & Hit@3 & Hit@10 & MRR & Hit@1 & Hit@3 & Hit@10 \\
\midrule
\multirow{3}{*}{KGC}   & TransE & 24.86 & 12.78 & 31.48 & 47.07 & 29.19 & 21.06 & 33.20 & 44.23 & 30.73 & 23.45 & 35.18 & 43.37 \\
                        & Tucker & 33.86 & 25.33 & 37.91 & 50.38 & 30.39 & 24.44 & 32.91 & 41.25 & 37.05 & 34.59 & 38.43 & 41.45 \\
                        & RotatE & 29.28 & 17.87 & 36.12 & 49.66 & 33.67 & 26.80 & 36.68 & 46.76 & 34.95 & 29.10 & 38.35 & 45.30 \\
\midrule
\multirow{10}{*}{MMKGC} & IKRL & 26.82 & 14.09 & 34.93 & 49.09 & 32.36 & 26.11 & 34.75 & 44.07 & 33.22 & 30.37 & 34.28 & 38.60 \\
                        & RSME & 29.80 & 24.20 & 32.10 & 49.40 & 29.20 & 23.40 & 32.00 & 40.40 & 34.40 & 33.80 & 36.10 & 38.60 \\
                        
                        & AdaMF & 32.51 & 21.31 & 39.67 & 51.68 & 34.27 & 27.21 & 37.86 & 47.21 & 38.06 & 33.49 & 40.44 & 45.48 \\
                        & OTKGE & 23.86 & 18.45 & 25.89 & 34.23 & 34.36 & 34.36 & 36.25 & 44.88 & 35.51 & 31.97 & 37.18 & 41.38 \\
                        & VISTA & 30.42 & 22.49 & 33.56 & 45.94 & 32.91 & 26.12 & 35.38 & 45.61 & 30.45 & 24.87 & 32.39 & 41.53 \\
                        & QEB & 28.18 & 14.82 & 36.67 & 51.55 & 32.38 & 25.47 & 35.06 & 45.32 & 34.37 & 29.49 & 36.95 & 42.32 \\
                        & IMF & 32.25 & 24.20 & 36.00 & 48.19 & 34.50 & 28.77 & 36.62 & 45.44 & 35.79 & 32.95 & 37.14 & 40.63 \\
                        & MMRNS & 32.68 & 23.01 & 37.86 & 51.01 & 35.03 & 28.59 & 37.49 & 47.47 & 35.93 & 30.53 & 39.07 & 45.47 \\
                        & SNAG$^\clubsuit$ & 36.30 & 27.40 & 41.10 & \underline{53.00} & \underline{37.30} & \underline{30.20} & \underline{40.50} & \underline{50.30} & 39.10 & 34.7 & 41.08 & \textbf{46.70} \\
                        & NativE$^\clubsuit$ & 34.30 & 25.08 & 39.48 & 51.35 & 36.84 & 29.94 & 40.06 & 49.39 & \underline{39.21} & \underline{35.03} & \underline{41.21} & 46.25 \\
                        & MyGO & \underline{37.72} & \underline{30.08} & \underline{41.26} & 52.21 & 36.10 & 29.78 & 38.54 & 47.75 & 38.44 & 35.01 & 39.84 & 44.19 \\
\midrule
\rowcolor[gray]{0.9} & \textbf{TSAM(Ours)} & \textbf{40.50} & \textbf{32.60} & \textbf{44.36} & \textbf{55.44} & \textbf{40.07} & \textbf{33.29} & \textbf{42.53} & \textbf{52.72} &  \textbf{39.80} &  \textbf{35.28} &  \textbf{41.29} & \underline{46.44} \\
\rowcolor[gray]{0.9} & \textbf{improvement} & \textit{\textbf{7.37\%}} & \textit{\textbf{8.38\%}} & \textit{\textbf{7.51\%}} & \textit{\textbf{4.60\%}} & \textit{\textbf{7.43\%}} & \textit{\textbf{10.23\%}} & \textit{\textbf{5.01\%}} & \textit{\textbf{4.81\%}} & \textit{\textbf{1.5\%}} & \textit{\textbf{0.7\%}} & \textit{\textbf{0.2\%}} & - \\
\bottomrule
\end{tabular}
\end{table*}


\subsubsection{Evaluation metrics}
TSAM uses four key evaluation metrics in the MMKGC task: mean reciprocal rank (MRR) and Hits@1, Hits@3, and Hits@10. The calculation of MRR and Hits@N is as follows:

\begin{small}
    \begin{gather}
        \mathbf{MRR} = \frac{1}{|E|} \sum_{i=1}^{|E|} \frac{1}{\operatorname{rank}(i)} = \frac{1}{|N|}\left(\frac{1}{\operatorname{rank}(1)} + \cdots + \frac{1}{\operatorname{rank}(|E|)}\right) \\
        \text{Hits@}N = \frac{1}{|E|} \sum_{i=1}^{|E|} \mathbb{I}\left(\operatorname{rank}_i \leq N\right)
    \end{gather}
\end{small}

where $|T|$ represents the number of triples in the set, and $rank_i$ represents the ranking position of the link prediction of the $i$th triple. ” And $\text{II}(\cdot)$ is a binary function that outputs a value of 1 if the judgment is true, otherwise it outputs a value of 0. In our experiments, $n=1,3,10$ is used.
\subsection{Baselines and Implementation Detail}
\subsubsection{Baselines}

To verify the effectiveness of the TSAM model, we selected 13 different types of methods as baseline models for comparison, including 3 classic single-modal baseline models: TransE\cite{bordes2013translating}, RotatE\cite{sun2019rotate}, Tucker\cite{balavzevic2019tucker}, as well as dozens of MMKGC models as demonstrated below:  IKRL \cite{xie2016image},  AdaMF \cite{zheng2024adapting}, OTKGE \cite{cao2022otkge}, VISTA\cite{lee2023vista}, RSME \cite{wang2021visual},  QEB \cite{wang2023tiva}, IMF \cite{li2023imf}, MMRNS \cite{xu2022relation},  MyGO \cite{zhang2024mygo}. SNAG \cite{chen2024power}. NativE \cite{zhang2024native}.

\subsubsection{Implementation Detail}
We use the pytorch \cite{paszke2019pytorch} framework to implement the TSAM model. For the text and visual modalities in the DB15K\cite{liu2019mmkg} and MKG-W\cite{xu2022relation} datasets, we follow our previous work \cite{zhang2024mygo} and use BEIT-V2\cite{peng2022beit} and BERT\cite{devlin2018bert} as tokenizers, respectively. We use bert-base as the main transformer encoder and decoder of the model, and use bert-large, RoBERTa-base/large\cite{liu2019roberta}, LLaMA-7B\cite{touvron2023llama} and DeBERTa-base/large\cite{he2020deberta} as variant models for cross-validation. TSAM uses the Adam\cite{kingma2014adam} optimizer to optimize model parameters. All experiments on TSAM were conducted on a Linux Ubuntu server equipped with 8 NVIDIA TESLA V100 32G GPUs. The code is available at https://github.com/2391134843/TSAM.

\subsection{Main Results}

According to the results in Table II, we can easily see the following situations:

1. Traditional models that only use a single mode, such as TransE, TuckER, and RotatE, usually show lower performance because they do not utilize multi-modal knowledge.

2. Models such as AdaMF, VISTA, IMF, and NativE outperform single-modality models by combining image and text modalities. For example, SNAG achieves 36.30\% MRR and 53.00\% Hit@10 on DB15K, and 37.30\% MRR and 50.30\% Hit@10 on MKG-W, highlighting the advantages of multi-modal knowledge.

3. The proposed TSAM model achieves the best performance on most metrics on all datasets, with an improvement of about 1\%-10\%. These results highlight the ability of TSAM to integrate fine-grained multi-modal information, align other modalities with the graph structure modality, and significantly improve prediction accuracy and overall performance. In addition, we found that the improvement ratio of the TSAM model for Hits@1 and MRR metrics is usually large, which means that the model achieves the best results in both overall prediction and accurate prediction.

\subsection{Ablation experiment}
To explore the contribution and impact of different model components on the model, we compared TSAM with the following three types of variants: (1) w/o FgMAF: a version without a fine-grained modality-aware fusion method. (2) w/o SaCL: a version without a structure-aware contrastive learning method. (3) w/0 $\mathcal{L}_{S T}$ a version without the structure-text contrastive loss.    (4) w/0 $\mathcal{L}_{S V}$ a version without the structure-visual contrastive loss. (5) TSAM-TransE/RotatE: a model that uses the TransE/RotatE model as a scoring function and obtains structural modality entity and relation embeddings. (6) Using different Decoder models to explore the trend of model effect changes.

\begin{table}[h]
\centering
\caption{Ablation experiments of TSAM on DB15K and MKG-W datasets.}

\begin{tabular}{lcccc}
\toprule
\multirow{2}{*}{} & \multicolumn{4}{c}{DB15K} \\ \cmidrule{2-5}
& MRR & Hit@1 & Hit@3 & Hit@10 \\ \midrule
\rowcolor[gray]{0.9} TSAM & \textbf{40.50} & \textbf{32.60} & \textbf{44.36} & \textbf{55.44} \\
w/o FgMAF & 39.83 & 31.90 & 43.73 & 54.55 \\
w/o SaCL & 38.83 & 31.11 & 42.87 & 53.95 \\ 
w/o $\mathcal{L}_{S T}$ & 39.33 & 31.62 & 42.99 & 54.31 \\ 
w/o $\mathcal{L}_{S V}$ & 39.37 & 31.65 & 42.93 & 54.36 \\ \midrule
\midrule
\multirow{2}{*}{} & \multicolumn{4}{c}{MKG-W} \\ \cmidrule{2-5}
& MRR & Hit@1 & Hit@3 & Hit@10 \\ \midrule
\rowcolor[gray]{0.9} TSAM & \textbf{40.07} & \textbf{33.29} & \textbf{42.53} & \textbf{52.72} \\
w/o FgMAF & 38.88 & 32.60 & 41.54 & 51.02 \\
w/o SaCL & 37.30 & 31.23 & 39.45 & 48.82 \\ 
w/o $\mathcal{L}_{S T}$ & 38.63 & 32.26 & 41.12 & 50.95 \\ 
w/o $\mathcal{L}_{S V}$ & 38.51 & 32.16 & 40.83 & 50.6 \\ \bottomrule
\end{tabular}%

\end{table}


\subsubsection{Analyze the Impact of FgMAF, SaCL, $\mathcal{L}_{S T}$ and $\mathcal{L}_{S V}$: } Analyze the impact of the FgMAF and SaCL The ablation results in Table III demonstrate the significant contributions of FgMAF and SaCL to the performance of TSAM. Removing FgMAF (w/o FgMAF) led to performance drops on DB15K by 1.65\% (MRR), 2.15\% (Hit@1), 1.42\% (Hit@3), and 1.60\% (Hit@10), and on MKG-W by 2.97\% (MRR), 2.07\% (Hit@1), 2.33\% (Hit@3), and 3.22\% (Hit@10). These results highlight the importance of fine-grained pre-trained models and attention mechanisms in enhancing the perceptual fusion of multi-modal semantic information. Similarly, removing SaCL (w/o SaCL) resulted in more significant performance declines: on DB15K, 4.12\% (MRR), 4.57\% (Hit@1), 3.36\% (Hit@3), and 2.69\% (Hit@10), and on MKG-W, 6.92\% (MRR), 6.18\% (Hit@1), 7.23\% (Hit@3), and 7.39\% (Hit@10). These findings validate that SaCL aligns other modalities effectively with the structural modality, reinforcing robust entity representations and being consistent with the previous viewpoint of this article. In addition, we also ablated the effects of two parts, namely the structure-text contrast loss $\mathcal{L}_{S T}$ and the structure-visual contrast loss $\mathcal{L}_{S V}$. Experimental results show that removing any loss will lead to a decrease in model performance, indicating that they play a key role in aligning the structure with other modalities and improving model performance.

\subsubsection{Analyze the impact of different scoring functions}
\begin{table}[h]
\centering
\caption{Experimental results of the TSAM model using different message functions on the MKG-W dataset.}
\begin{tabular}{lcccc}
\toprule
Model & MRR & Hit@1 & Hit@3 & Hit@10 \\
\midrule
TSAM-Tucker & \textbf{40.07} & \textbf{33.29} & \textbf{42.53} & \textbf{52.72} \\
TSAM-RotateE & 37.49 & 31.09 & 39.67 & 49.46 \\
TSAM-TransE & 36.77 & 30.59 & 38.74 & 48.29 \\
\bottomrule
\end{tabular}
\end{table}
From Table IV, it can be seen that TSAM-Trucker performs the best overall, while TSAM-TranE shows relatively weaker performance. These results indicate that the choice of scoring function significantly impacts the performance of the MMKGC model. In practical applications, selecting an appropriate scoring function should be balanced according to the specific requirements of the task.

Through ablation experiments, it can be observed that both the FgMAF and SaCL methods in the TSAM model contribute significantly to enhancing model performance. The FgMAF method effectively improves the model's ability to interact with and perceive fine-grained modal information, while the SaCL method plays a crucial role in alignment, primarily guided by structural modal knowledge. However, in terms of the proportion of improvement, SaCL surpasses FgMAF, which supports the initial hypothesis of this study: graph-structured modal knowledge is the most critical for the MMKGC task, and other modalities should align with the structural modality. Overall, the TSAM model demonstrates excellent performance in the multi-modal knowledge graph completion task, validating its design's rationale and effectiveness.


\subsubsection{Experiments on different Transformer-based Decoder models}
\begin{figure*}[t!] 
\centering
\includegraphics[width=\textwidth]{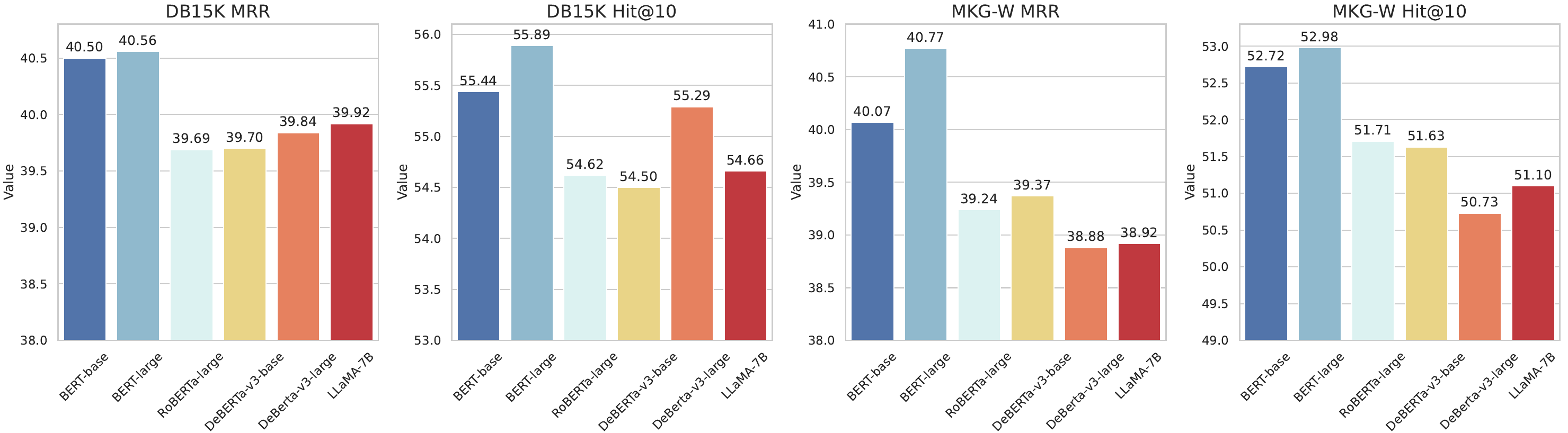}
\caption{The experimental results of the TSAM model using Bert-base/large, RoBERTa- large, and DeBERTa-base/large as decoders on the DB15K and MKG-W datasets.}
\label{figure4-1}
\end{figure*}
As an important part of TSAM, we studied the impact of using different decoder models. From Fig.4, we can see three interesting phenomena: (1) BERT-large outperforms other models on both datasets, especially in terms of accurate prediction (Hit@1) and overall prediction ability (MRR). (2) For the same type of model, the large version with more parameters tends to perform better than the basic model. (3) The effect of the large prediction model LLaMA-7B\cite{touvron2023llama} on the MMKGC task did not achieve the expected effect, and it was even worse than most pre-trained language models.

\begin{figure}[ht!] 
\centering
\includegraphics[width=\columnwidth]{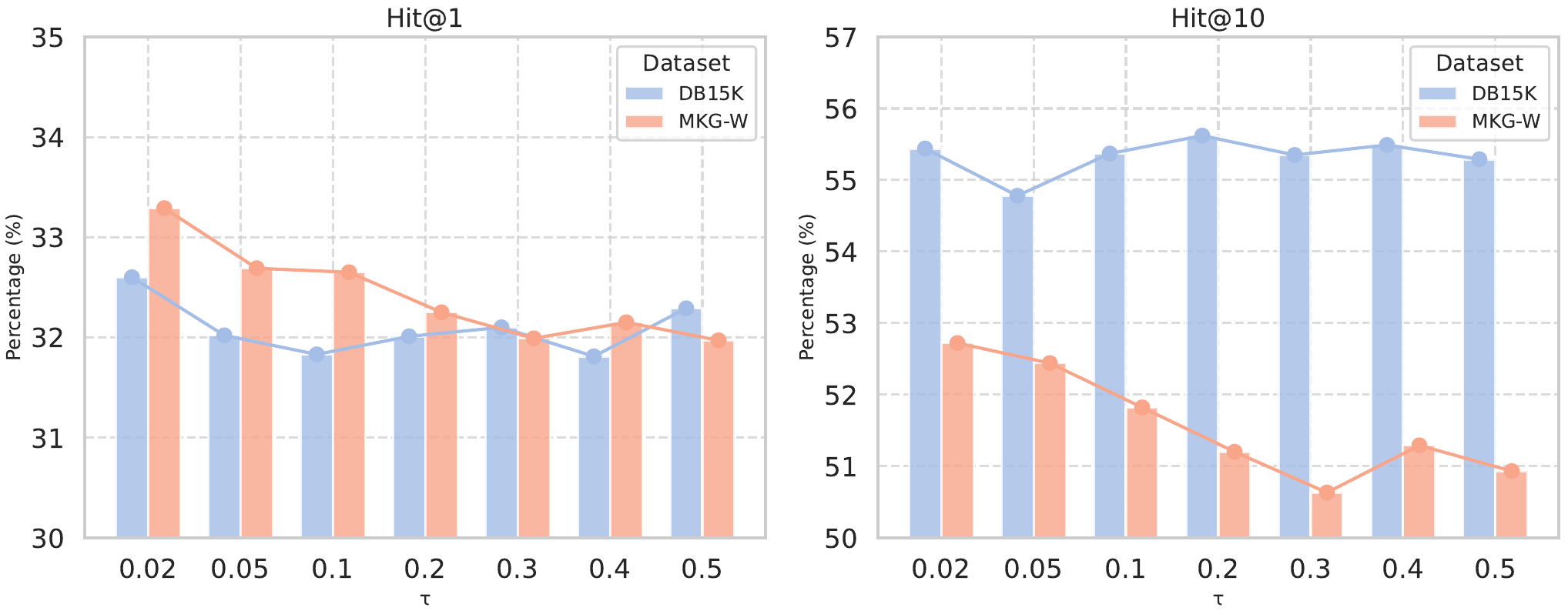}
\caption{Parameter sensitivity experiment of the number of Temperature parameter $\tau$}
\label{figure5}
\end{figure}

\subsection{Parameter sensitivity experiments}

In order to further explore the sensitivity of the model to important parameters in contrastive learning, we studied the effects of the contrastive learning temperature parameter $\tau$ and the number of negative samples K in the batch on the model.

\subsubsection{Temperature parameter $\tau$}The experimental results of the TSAM model on the two datasets are shown in Fig.5. Smaller $\tau$ values (such as 0.02) perform best on the DB15K and MKG-W datasets. This shows that in the TSAM model, smaller $\tau$ values help to enhance the effect of contrastive learning between modalities, thereby more effectively alleviating the semantic gap between modalities. When $\tau$ increases, the model performance decreases to a small extent, especially in the Hit@10 indicator. Let's make a simple analysis from the principle behind contrastive learning\cite{wang2021understanding}. When $\tau$ is small, the similarity score is amplified, the similarity score of the positive sample will be relatively higher, and the similarity score of the negative sample will be relatively lower. When $\tau$ is large, the similarity score is reduced, and the similarity scores of the positive and negative samples will be closer. Formally speaking, as $\tau$ decreases, other modalities will be closer to the structural modality. The emergence of this phenomenon is also consistent with the starting point of our paper, confirming the importance of structural modality.

\begin{table}[ht]
\centering
\caption{Parameter sensitivity experiment of the number of negative samples K}
\begin{tabular}{@{}lccccc@{}}
\toprule
\multirow{2}{*}{Neg\_Num} & \multicolumn{5}{c}{DB15K} \\ 
\cmidrule(lr){2-6}
                         & MRR   & Hit@1 & Hit@3 & Hit@10 & Mem. Usage \\ \midrule
K=8                      & 39.50  & 31.78 & 43.03 & 54.65  & 16.77G     \\
K=16                     & \textbf{40.50}  & \textbf{32.60}  & 44.36 & 55.44  & 17.47G     \\
K=32                     & 40.07 & 32.36 & 43.44 & 55.00  & 20.24G     \\
K=64                     & 40.16 & 32.24 & \textbf{43.93} & \textbf{55.59} & 20.99G     \\ 
\midrule 
\midrule
\multirow{2}{*}{Neg\_Num} & \multicolumn{5}{c}{MKG-W} \\ 
\cmidrule(lr){2-6}
                         & MRR   & Hit@1 & Hit@3 & Hit@10 & Mem. Usage \\ \midrule
K=8                      & 39.80  & 33.27 & 42.23 & 52.24  & 19.26G     \\
K=16                     & 40.07 & 33.29 & 42.53 & \textbf{52.72}  & 19.96G     \\
K=32                     & 40.10  & 33.29 & 42.70  & 52.72  & 22.55G     \\
K=64                     & \textbf{40.20}  & \textbf{33.37} & \textbf{43.05} & 52.71  & 24.23G     \\ \bottomrule
\end{tabular}
\label{tab:performance}
\end{table}
\subsubsection{The number of negative samples K} The experimental results in Table V show that using 16 negative samples K achieves almost the best MRR and Hit@1 performance on DB15K and MKG-W datasets. Although increasing K to 32 or 64 slightly improves the performance, the gain is minimal and the server resource consumption increases significantly. More negative samples can stabilize representation learning by better distinguishing adversarial noise and enhancing modality discrimination. However, too many negative samples increase the computational overhead and the performance improvement weakens as the negative pairs become more similar. Therefore, a moderate amount of negative samples is the best choice to balance performance and efficiency.

\subsection{Experiments with different numbers of modality tokens} 

\begin{figure}[ht!] 
\centering
\includegraphics[width=3.5in]{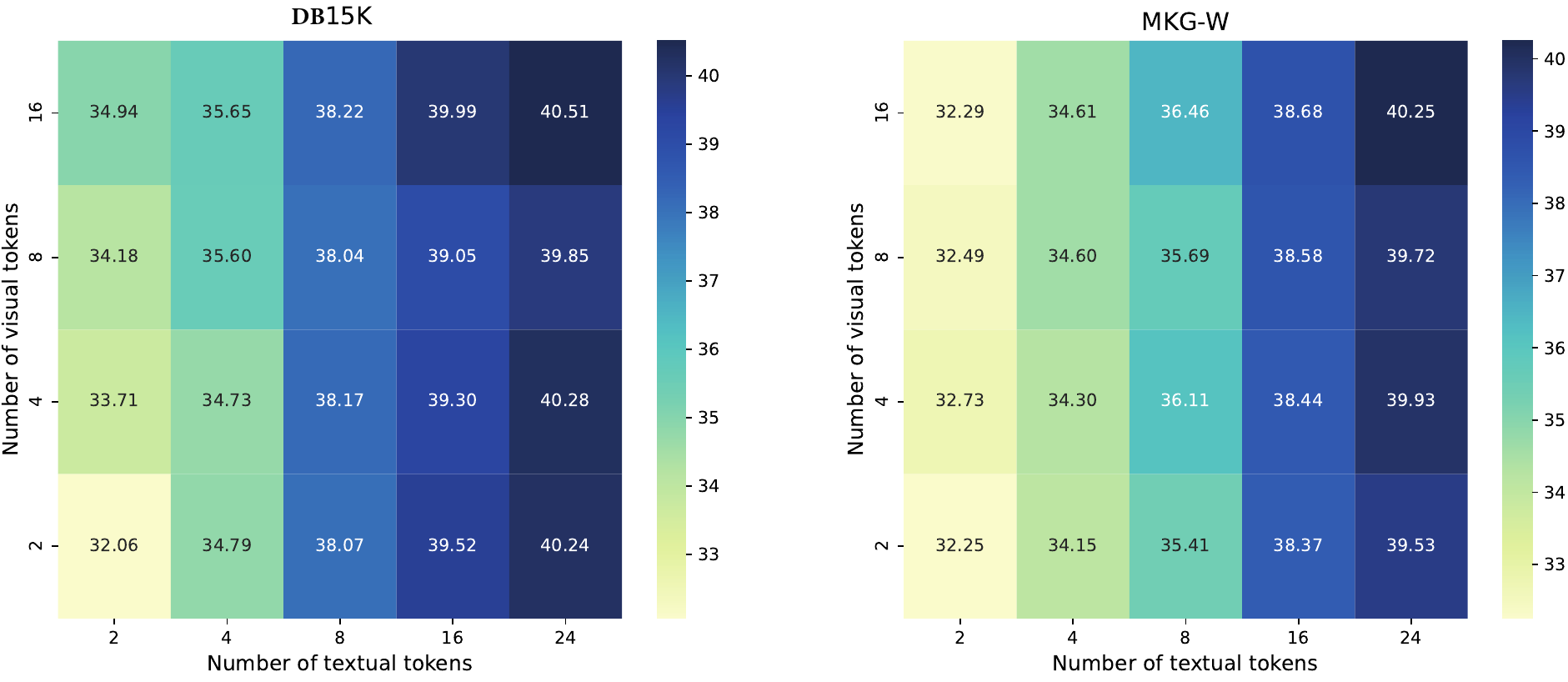}
\caption{The MRR performance of the TSAM model on the DB15K and MKG-W datasets with varying numbers of modality tokens.}
\label{figure4}
\end{figure}

We analyzed the impact of multi-modal token quantities on TSAM's performance using experiments on DB15K and MKG-W, as shown in Fig.6. Results indicate that increasing visual and textual tokens enhances the model's MRR by enriching entity and relation features through complementary multi-modal information. A balanced increase in both modalities yields the most significant gains, emphasizing the importance of multi-modal fusion. However, performance improvement slows at higher token levels (e.g., 16 or 24), likely due to feature redundancy and computational overhead. Practical applications should balance token quantity and efficiency.

\subsection{Case Study}
\begin{table}[]
\centering

\caption{Case study experiments on the MKG-W dataset}
\begin{tabular}{l|l}
\toprule
\multicolumn{2}{c}{\textbf{Case 1}: ($J.R.R. Tolkien, ethnic\ group, ?$)} \\ \midrule
\rowcolor[gray]{0.9} \textbf{TSAM} & \textbf{Rank} of the correct tail entity $"English\ people"$ : \underline{1}     \\ \midrule
\textbf{MyGo} & \textbf{Rank} of the correct tail entity $"English\ people"$ : \underline{2}      \\ \bottomrule 
\multicolumn{2}{l}{\vspace{-0.8em}}  \\ \toprule
\multicolumn{2}{c}{\textbf{Case 2}: ($Oasis , influenced\ by, ?$)}          \\ \midrule
\rowcolor[gray]{0.9} \textbf{TSAM} & \textbf{Rank} of the correct tail entity $"The\ Beatles"$ : \underline{1}         \\ \midrule
\textbf{MyGo} & \textbf{Rank} of the correct tail entity $"The\ Beatles"$ : \underline{9}         \\ \bottomrule
\multicolumn{2}{l}{\vspace{-0.8em}}  \\ \toprule
\multicolumn{2}{c}{\textbf{Case 3}: ($Son\ of\ Paleface, cast\ member, ?$)}    \\ \midrule
\rowcolor[gray]{0.9} \textbf{TSAM} & \textbf{Rank} of the correct tail entity $"Bing\ Crosby"$ : \underline{1}         \\ \midrule
\textbf{MyGo} & \textbf{Rank} of the correct tail entity $"Bing\ Crosby"$ : \underline{36}        \\ \bottomrule
\multicolumn{2}{l}{\vspace{-0.8em}}  \\ \toprule
\multicolumn{2}{c}{\textbf{Case 4}: ($Paris\ Underground, director, ?$)}     \\ \midrule
\rowcolor[gray]{0.9} \textbf{TSAM} & \textbf{Rank} of the correct tail entity $"Gregory\ Ratoff"$ : \underline{1}      \\ \midrule
\textbf{MyGo} & \textbf{Rank} of the correct tail entity $"Gregory\ Ratoff"$ : \underline{5119}   \\ \bottomrule
\end{tabular}
\end{table}
As shown in Table VI, we selected several representative cases from a batch for analysis, examining the specific performance of each triple to verify the effectiveness of TSAM from the most intuitive perspective. It is evident that the TSAM model is better at capturing simple structured information when handling triples compared to the MyGo model. This is particularly true in cases where MyGo\cite{zhang2024mygo} fails to answer structurally strong triples (where the answer is very fixed and singular), yet TSAM can still predict the correct result. For example, when predicting the triple \textit{(Paris Underground, director, ?)}, MyGO is completely unable to provide the correct answer, while TSAM continues to accurately predict the correct answer. These findings demonstrate that the TSAM model, with its finer-grained modality capture and greater emphasis on graph structure, effectively reduces noise from other modalities on the structural modality. This leads to improved entity representations and, consequently, better MMKGC performance.

\subsection{Visualization}
\begin{figure}[ht!] 
\centering
\includegraphics[width=3.5in]{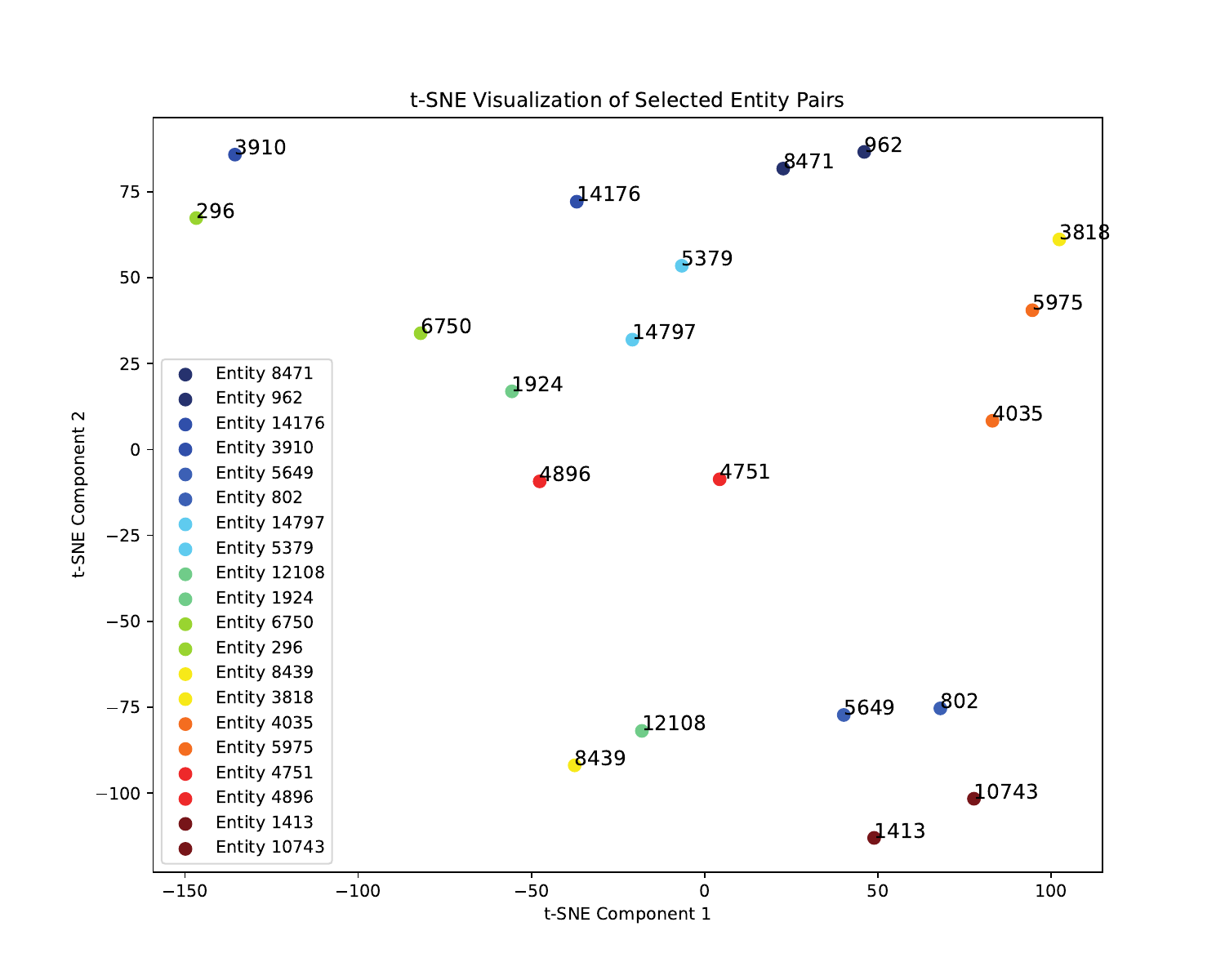}
\caption{Use t-SNE\cite{van2008visualizing} to visualize the dimensionality reduction of triplet embeddings in small batch size.}
\label{figure4}
\end{figure}
In this experiment, points of the same color represent the head and tail entities of the same triple. Fig.7 clearly demonstrates that the TSAM model effectively places the embeddings of the same triple in close proximity. For example, the triples (1413, 10743) and (5649, 802) represent \textit{"Dev is directed by Govind Nihalani"} and \textit{"The 1954 film Thookku Thookki starred actor Sivaji Ganesan "} respectively\footnote{Detailed mapping is available at: https://github.com/2391134843/TSAM}. This visualization highlights TSAM's capability to accurately cluster the embeddings of related entities, reflecting the semantic relationships within each triple.

Careful consideration of the mapping of the triples reveals that these entities related to Indian movies are closer in space after dimensionality reduction. It is worth noting that the triples do not clearly indicate that these entities belong to India, but the model embeds these entities in a closer range through multi-modal semantic learning. 
This demonstrates that the model can more effectively capture the fine-grained modal knowledge perception and interactions between entities and relations in MMKG. By integrating modal-aware contrastive learning, the model enhances the learning of their potential feature representations, thereby significantly improving the performance of the MMKGC model.

\section{Conclusion And Future Work}

In this paper, we propose TSAM, a novel model for multi-modal knowledge graph completion. TSAM addresses two key challenges: (1) a fine-grained modality-aware fusion method that captures and integrates semantic information across modalities using pre-trained models and attention mechanisms, and (2) a structure-aware contrastive learning approach that aligns modalities to the structural modality, reducing noise during fusion. Experimental results on three benchmarks show that TSAM significantly outperforms existing models, highlighting its effectiveness and offering new insights for multi-modal knowledge graph completion research.

Although TSAM has achieved significant performance improvements, there are still some things we have not yet completed. For example, the current model's ability to be applied to large-scale dynamic knowledge graphs has not been fully verified. Future plans include the following aspects: 1) Explore the synergy between TSAM and pre-trained language models to further enhance the semantic understanding of text modality; 2) Develop an incremental structural contrast learning framework to adapt to the dynamic update characteristics of knowledge graphs; 3) Construct a fine-grained modality credibility evaluation indicator to achieve a smarter modality weight allocation. 4) Future work will test TSAM on a larger MMKGC dataset to further verify its scalability.


\bibliography{references} 

\begin{thebibliography}{10}
\providecommand{\url}[1]{#1}
\csname url@samestyle\endcsname
\providecommand{\newblock}{\relax}
\providecommand{\bibinfo}[2]{#2}
\providecommand{\BIBentrySTDinterwordspacing}{\spaceskip=0pt\relax}
\providecommand{\BIBentryALTinterwordstretchfactor}{4}
\providecommand{\BIBentryALTinterwordspacing}{\spaceskip=\fontdimen2\font plus
\BIBentryALTinterwordstretchfactor\fontdimen3\font minus
  \fontdimen4\font\relax}
\providecommand{\BIBforeignlanguage}[2]{{%
\expandafter\ifx\csname l@#1\endcsname\relax
\typeout{** WARNING: IEEEtran.bst: No hyphenation pattern has been}%
\typeout{** loaded for the language `#1'. Using the pattern for}%
\typeout{** the default language instead.}%
\else
\language=\csname l@#1\endcsname
\fi
#2}}
\providecommand{\BIBdecl}{\relax}
\BIBdecl

\bibitem{liang2024survey}
K.~Liang, L.~Meng, M.~Liu, Y.~Liu, W.~Tu, S.~Wang, S.~Zhou, X.~Liu, F.~Sun, and
  K.~He, ``A survey of knowledge graph reasoning on graph types: Static,
  dynamic, and multi-modal,'' \emph{IEEE Transactions on Pattern Analysis and
  Machine Intelligence}, 2024.

\bibitem{ji2021survey}
S.~Ji, S.~Pan, E.~Cambria, P.~Marttinen, and S.~Y. Philip, ``A survey on
  knowledge graphs: Representation, acquisition, and applications,'' \emph{IEEE
  transactions on neural networks and learning systems}, vol.~33, no.~2, pp.
  494--514, 2021.

\bibitem{yi2021multi}
J.~Yi and Z.~Chen, ``Multi-modal variational graph auto-encoder for
  recommendation systems,'' \emph{IEEE Transactions on Multimedia}, vol.~24,
  pp. 1067--1079, 2021.

\bibitem{zheng2024adapting}
B.~Zheng, Y.~Hou, H.~Lu, Y.~Chen, W.~X. Zhao, M.~Chen, and J.-R. Wen,
  ``Adapting large language models by integrating collaborative semantics for
  recommendation,'' in \emph{2024 IEEE 40th International Conference on Data
  Engineering (ICDE)}.\hskip 1em plus 0.5em minus 0.4em\relax IEEE, 2024, pp.
  1435--1448.

\bibitem{wu2022graph}
S.~Wu, F.~Sun, W.~Zhang, X.~Xie, and B.~Cui, ``Graph neural networks in
  recommender systems: a survey,'' \emph{ACM Computing Surveys}, vol.~55,
  no.~5, pp. 1--37, 2022.

\bibitem{cao2022cross}
X.~Cao, Y.~Shi, J.~Wang, H.~Yu, X.~Wang, and Z.~Yan, ``Cross-modal knowledge
  graph contrastive learning for machine learning method recommendation,'' in
  \emph{Proceedings of the 30th ACM international conference on multimedia},
  2022, pp. 3694--3702.

\bibitem{yang2023knowledge}
Y.~Yang, C.~Huang, L.~Xia, and C.~Huang, ``Knowledge graph self-supervised
  rationalization for recommendation,'' in \emph{Proceedings of the 29th ACM
  SIGKDD conference on knowledge discovery and data mining}, 2023, pp.
  3046--3056.

\bibitem{cao2022building}
L.~Cao, H.~Zhang, and L.~Feng, ``Building and using personal knowledge graph to
  improve suicidal ideation detection on social media,'' \emph{IEEE
  Transactions on Multimedia}, vol.~24, pp. 87--102, 2022.

\bibitem{yang2023context}
A.~Yang, S.~Lin, C.-H. Yeh, M.~Shu, Y.~Yang, and X.~Chang, ``Context matters:
  Distilling knowledge graph for enhanced object detection,'' \emph{IEEE
  Transactions on Multimedia}, vol.~26, pp. 487--500, 2023.

\bibitem{pan2024unifying}
S.~Pan, L.~Luo, Y.~Wang, C.~Chen, J.~Wang, and X.~Wu, ``Unifying large language
  models and knowledge graphs: A roadmap,'' \emph{IEEE Transactions on
  Knowledge and Data Engineering}, 2024.

\bibitem{chen2024knowledge}
Z.~Chen, Y.~Zhang, Y.~Fang, Y.~Geng, L.~Guo, X.~Chen, Q.~Li, W.~Zhang, J.~Chen,
  Y.~Zhu \emph{et~al.}, ``Knowledge graphs meet multi-modal learning: A
  comprehensive survey,'' \emph{arXiv preprint arXiv:2402.05391}, 2024.

\bibitem{ni2023psnea}
W.~Ni, Q.~Xu, Y.~Jiang, Z.~Cao, X.~Cao, and Q.~Huang, ``Psnea: Pseudo-siamese
  network for entity alignment between multi-modal knowledge graphs,'' in
  \emph{Proceedings of the 31st ACM International Conference on Multimedia},
  2023, pp. 3489--3497.

\bibitem{zhu2022multi}
X.~Zhu, Z.~Li, X.~Wang, X.~Jiang, P.~Sun, X.~Wang, Y.~Xiao, and N.~J. Yuan,
  ``Multi-modal knowledge graph construction and application: A survey,''
  \emph{IEEE Transactions on Knowledge and Data Engineering}, vol.~36, no.~2,
  pp. 715--735, 2022.

\bibitem{bordes2013translating}
A.~Bordes, N.~Usunier, A.~Garcia-Duran, J.~Weston, and O.~Yakhnenko,
  ``Translating embeddings for modeling multi-relational data,'' \emph{Advances
  in neural information processing systems}, vol.~26, 2013.

\bibitem{sun2019rotate}
Z.~Sun, Z.-H. Deng, J.-Y. Nie, and J.~Tang, ``Rotate: Knowledge graph embedding
  by relational rotation in complex space,'' \emph{arXiv preprint
  arXiv:1902.10197}, 2019.

\bibitem{cao-etal-2021-missing}
\BIBentryALTinterwordspacing
Y.~Cao, X.~Ji, X.~Lv, J.~Li, Y.~Wen, and H.~Zhang, ``Are missing links
  predictable? an inferential benchmark for knowledge graph completion,''
  C.~Zong, F.~Xia, W.~Li, and R.~Navigli, Eds.\hskip 1em plus 0.5em minus
  0.4em\relax Online: Association for Computational Linguistics, Aug. 2021, pp.
  6855--6865. [Online]. Available:
  \url{https://aclanthology.org/2021.acl-long.534/}
\BIBentrySTDinterwordspacing

\bibitem{wang2022simkgc}
L.~Wang, W.~Zhao, Z.~Wei, and J.~Liu, ``Simkgc: Simple contrastive knowledge
  graph completion with pre-trained language models,'' \emph{arXiv preprint
  arXiv:2203.02167}, 2022.

\bibitem{zeng2023matching}
W.~Zeng, X.~Zhao, Z.~Tan, J.~Tang, and X.~Cheng, ``Matching knowledge graphs in
  entity embedding spaces: an experimental study,'' \emph{IEEE Transactions on
  Knowledge and Data Engineering}, vol.~35, no.~12, pp. 12\,770--12\,784, 2023.

\bibitem{wei2024multi}
Y.~Wei, W.~Chen, X.~Zhang, P.~Zhao, J.~Qu, and L.~Zhao, ``Multi-modal siamese
  network for few-shot knowledge graph completion,'' in \emph{2024 IEEE 40th
  International Conference on Data Engineering (ICDE)}.\hskip 1em plus 0.5em
  minus 0.4em\relax IEEE, 2024, pp. 719--732.

\bibitem{liang2024mgksite}
K.~Liang, L.~Meng, H.~Li, M.~Liu, S.~Wang, S.~Zhou, X.~Liu, and K.~He,
  ``Mgksite: Multi-modal knowledge-driven site selection via intra and
  inter-modal graph fusion,'' \emph{IEEE Transactions on Multimedia}, 2024.

\bibitem{zheng2023mmkgr}
S.~Zheng, W.~Wang, J.~Qu, H.~Yin, W.~Chen, and L.~Zhao, ``Mmkgr: Multi-hop
  multi-modal knowledge graph reasoning,'' in \emph{2023 IEEE 39th
  International Conference on Data Engineering (ICDE)}.\hskip 1em plus 0.5em
  minus 0.4em\relax IEEE, 2023, pp. 96--109.

\bibitem{zhang2022multimodal}
N.~Zhang, L.~Li, X.~Chen, X.~Liang, S.~Deng, and H.~Chen, ``Multimodal
  analogical reasoning over knowledge graphs,'' in \emph{The Eleventh
  International Conference on Learning Representations}, 2022.

\bibitem{zhang2024native}
Y.~Zhang, Z.~Chen, L.~Guo, Y.~Xu, B.~Hu, Z.~Liu, W.~Zhang, and H.~Chen,
  ``Native: Multi-modal knowledge graph completion in the wild,'' in
  \emph{Proceedings of the 47th International ACM SIGIR Conference on Research
  and Development in Information Retrieval}, 2024, pp. 91--101.

\bibitem{zhao2024contrast}
Y.~Zhao, Y.~Zhang, B.~Zhou, X.~Qian, K.~Song, and X.~Cai, ``Contrast then
  memorize: Semantic neighbor retrieval-enhanced inductive multimodal knowledge
  graph completion,'' in \emph{Proceedings of the 47th International ACM SIGIR
  Conference on Research and Development in Information Retrieval}, 2024, pp.
  102--111.

\bibitem{zhang2024mygo}
Y.~Zhang, Z.~Chen, L.~Guo, Y.~Xu, B.~Hu, Z.~Liu, H.~Chen, and W.~Zhang, ``Mygo:
  Discrete modality information as fine-grained tokens for multi-modal
  knowledge graph completion,'' \emph{arXiv preprint arXiv:2404.09468}, 2024.

\bibitem{radford2021learning}
A.~Radford, J.~W. Kim, C.~Hallacy, A.~Ramesh, G.~Goh, S.~Agarwal, G.~Sastry,
  A.~Askell, P.~Mishkin, J.~Clark \emph{et~al.}, ``Learning transferable visual
  models from natural language supervision,'' in \emph{International conference
  on machine learning}.\hskip 1em plus 0.5em minus 0.4em\relax PMLR, 2021, pp.
  8748--8763.

\bibitem{peng2022beit}
Z.~Peng, L.~Dong, H.~Bao, Q.~Ye, and F.~Wei, ``Beit v2: Masked image modeling
  with vector-quantized visual tokenizers,'' \emph{arXiv preprint
  arXiv:2208.06366}, 2022.

\bibitem{zhu2023minigpt}
D.~Zhu, J.~Chen, X.~Shen, X.~Li, and M.~Elhoseiny, ``Minigpt-4: Enhancing
  vision-language understanding with advanced large language models,''
  \emph{arXiv preprint arXiv:2304.10592}, 2023.

\bibitem{bao2021beit}
H.~Bao, L.~Dong, S.~Piao, and F.~Wei, ``Beit: Bert pre-training of image
  transformers,'' \emph{arXiv preprint arXiv:2106.08254}, 2021.

\bibitem{devlin2018bert}
J.~Devlin, ``Bert: Pre-training of deep bidirectional transformers for language
  understanding,'' \emph{arXiv preprint arXiv:1810.04805}, 2018.

\bibitem{liu2019roberta}
Y.~Liu, ``Roberta: A robustly optimized bert pretraining approach,''
  \emph{arXiv preprint arXiv:1907.11692}, vol. 364, 2019.

\bibitem{he2020deberta}
P.~He, X.~Liu, J.~Gao, and W.~Chen, ``Deberta: Decoding-enhanced bert with
  disentangled attention,'' \emph{arXiv preprint arXiv:2006.03654}, 2020.

\bibitem{vaswani2017attention}
A.~Vaswani, ``Attention is all you need,'' \emph{Advances in Neural Information
  Processing Systems}, 2017.

\bibitem{balavzevic2019tucker}
I.~Bala{\v{z}}evi{\'c}, C.~Allen, and T.~M. Hospedales, ``Tucker: Tensor
  factorization for knowledge graph completion,'' \emph{arXiv preprint
  arXiv:1901.09590}, 2019.

\bibitem{wang2021structure}
B.~Wang, T.~Shen, G.~Long, T.~Zhou, Y.~Wang, and Y.~Chang,
  ``Structure-augmented text representation learning for efficient knowledge
  graph completion,'' in \emph{Proceedings of the Web Conference 2021}, 2021,
  pp. 1737--1748.

\bibitem{yao2019kg}
L.~Yao, C.~Mao, and Y.~Luo, ``Kg-bert: Bert for knowledge graph completion,''
  \emph{arXiv preprint arXiv:1909.03193}, 2019.

\bibitem{zhu2021neural}
Z.~Zhu, Z.~Zhang, L.-P. Xhonneux, and J.~Tang, ``Neural bellman-ford networks:
  A general graph neural network framework for link prediction,''
  \emph{Advances in Neural Information Processing Systems}, vol.~34, pp.
  29\,476--29\,490, 2021.

\bibitem{vashishth2019composition}
S.~Vashishth, S.~Sanyal, V.~Nitin, and P.~Talukdar, ``Composition-based
  multi-relational graph convolutional networks,'' \emph{arXiv preprint
  arXiv:1911.03082}, 2019.

\bibitem{li2022knowledge}
L.~Li, X.~Zhang, Y.~Ma, C.~Gao, J.~Wang, Y.~Yu, Z.~Yuan, and Q.~Ma, ``A
  knowledge graph completion model based on contrastive learning and relation
  enhancement method,'' \emph{Knowledge-Based Systems}, vol. 256, p. 109889,
  2022.

\bibitem{geng2023relational}
Y.~Geng, J.~Chen, J.~Z. Pan, M.~Chen, S.~Jiang, W.~Zhang, and H.~Chen,
  ``Relational message passing for fully inductive knowledge graph
  completion,'' in \emph{2023 IEEE 39th International Conference on Data
  Engineering (ICDE)}.\hskip 1em plus 0.5em minus 0.4em\relax IEEE, 2023, pp.
  1221--1233.

\bibitem{liang2024survey2}
W.~Liang, P.~D. Meo, Y.~Tang, and J.~Zhu, ``A survey of multi-modal knowledge
  graphs: Technologies and trends,'' \emph{ACM Computing Surveys}, vol.~56,
  no.~11, pp. 1--41, 2024.

\bibitem{cao2022otkge}
Z.~Cao, Q.~Xu, Z.~Yang, Y.~He, X.~Cao, and Q.~Huang, ``Otkge: Multi-modal
  knowledge graph embeddings via optimal transport,'' \emph{Advances in Neural
  Information Processing Systems}, vol.~35, pp. 39\,090--39\,102, 2022.

\bibitem{shang2024lafa}
B.~Shang, Y.~Zhao, J.~Liu, and D.~Wang, ``Lafa: Multimodal knowledge graph
  completion with link aware fusion and aggregation,'' in \emph{Proceedings of
  the AAAI Conference on Artificial Intelligence}, vol.~38, no.~8, 2024, pp.
  8957--8965.

\bibitem{lee2024multimodal}
J.~Lee, Y.~Wang, J.~Li, and M.~Zhang, ``Multimodal reasoning with multimodal
  knowledge graph,'' \emph{arXiv preprint arXiv:2406.02030}, 2024.

\bibitem{li2023imf}
X.~Li, X.~Zhao, J.~Xu, Y.~Zhang, and C.~Xing, ``Imf: interactive multimodal
  fusion model for link prediction,'' in \emph{Proceedings of the ACM Web
  Conference 2023}, 2023, pp. 2572--2580.

\bibitem{zhang2022rethinking}
Z.~Zhang, J.~Wang, J.~Ye, and F.~Wu, ``Rethinking graph convolutional networks
  in knowledge graph completion,'' in \emph{Proceedings of the ACM Web
  Conference 2022}, 2022, pp. 798--807.

\bibitem{xie2016image}
R.~Xie, Z.~Liu, H.~Luan, and M.~Sun, ``Image-embodied knowledge representation
  learning,'' \emph{arXiv preprint arXiv:1609.07028}, 2016.

\bibitem{zhang2024unleashing}
Y.~Zhang, Z.~Chen, L.~Liang, H.~Chen, and W.~Zhang, ``Unleashing the power of
  imbalanced modality information for multi-modal knowledge graph completion,''
  \emph{arXiv preprint arXiv:2402.15444}, 2024.

\bibitem{lee2023vista}
J.~Lee, C.~Chung, H.~Lee, S.~Jo, and J.~Whang, ``Vista: Visual-textual
  knowledge graph representation learning,'' in \emph{Findings of the
  Association for Computational Linguistics: EMNLP 2023}, 2023, pp. 7314--7328.

\bibitem{wang2023tiva}
X.~Wang, B.~Meng, H.~Chen, Y.~Meng, K.~Lv, and W.~Zhu, ``Tiva-kg: A multimodal
  knowledge graph with text, image, video and audio,'' in \emph{Proceedings of
  the 31st ACM International Conference on Multimedia}, 2023, pp. 2391--2399.

\bibitem{xu2022relation}
D.~Xu, T.~Xu, S.~Wu, J.~Zhou, and E.~Chen, ``Relation-enhanced negative
  sampling for multimodal knowledge graph completion,'' in \emph{Proceedings of
  the 30th ACM international conference on multimedia}, 2022, pp. 3857--3866.

\bibitem{paszke2019pytorch}
A.~Paszke, S.~Gross, F.~Massa, A.~Lerer, J.~Bradbury, G.~Chanan, T.~Killeen,
  Z.~Lin, N.~Gimelshein, L.~Antiga \emph{et~al.}, ``Pytorch: An imperative
  style, high-performance deep learning library,'' \emph{Advances in neural
  information processing systems}, vol.~32, 2019.

\bibitem{kingma2014adam}
D.~P. Kingma, ``Adam: A method for stochastic optimization,'' \emph{arXiv
  preprint arXiv:1412.6980}, 2014.

\bibitem{chen2020simple}
T.~Chen, S.~Kornblith, M.~Norouzi, and G.~Hinton, ``A simple framework for
  contrastive learning of visual representations,'' in \emph{International
  conference on machine learning}.\hskip 1em plus 0.5em minus 0.4em\relax PMLR,
  2020, pp. 1597--1607.

\bibitem{he2020mocov1}
K.~He, H.~Fan, Y.~Wu, S.~Xie, and R.~Girshick, ``Mocov1: Momentum contrast for
  unsupervised visual representation learning,'' 2020.

\bibitem{gao2021simcse}
T.~Gao, X.~Yao, and D.~Chen, ``Simcse: Simple contrastive learning of sentence
  embeddings,'' \emph{arXiv preprint arXiv:2104.08821}, 2021.

\bibitem{liu2019mmkg}
Y.~Liu, H.~Li, A.~Garcia-Duran, M.~Niepert, D.~Onoro-Rubio, and D.~S.
  Rosenblum, ``Mmkg: multi-modal knowledge graphs,'' in \emph{The Semantic Web:
  16th International Conference, ESWC 2019, Portoro{\v{z}}, Slovenia, June
  2--6, 2019, Proceedings 16}.\hskip 1em plus 0.5em minus 0.4em\relax Springer,
  2019, pp. 459--474.

\bibitem{ngiam2011multimodal}
J.~Ngiam, A.~Khosla, M.~Kim, J.~Nam, H.~Lee, and A.~Y. Ng, ``Multimodal deep
  learning,'' in \emph{Proceedings of the 28th international conference on
  machine learning (ICML-11)}, 2011, pp. 689--696.

\bibitem{yang2015embedding}
B.~Yang, S.~W.-t. Yih, X.~He, J.~Gao, and L.~Deng, ``Embedding entities and
  relations for learning and inference in knowledge bases,'' in
  \emph{Proceedings of the International Conference on Learning Representations
  (ICLR) 2015}, 2015.

\bibitem{lin2015learning}
Y.~Lin, Z.~Liu, M.~Sun, Y.~Liu, and X.~Zhu, ``Learning entity and relation
  embeddings for knowledge graph completion,'' in \emph{Proceedings of the AAAI
  conference on artificial intelligence}, vol.~29, no.~1, 2015.

\bibitem{gao2024embracing}
Z.~Gao, X.~Jiang, X.~Xu, F.~Shen, Y.~Li, and H.~T. Shen, ``Embracing unimodal
  aleatoric uncertainty for robust multimodal fusion,'' in \emph{Proceedings of
  the IEEE/CVF Conference on Computer Vision and Pattern Recognition}, 2024,
  pp. 26\,876--26\,885.

\bibitem{van2008visualizing}
L.~Van~der Maaten and G.~Hinton, ``Visualizing data using t-sne.''
  \emph{Journal of machine learning research}, vol.~9, no.~11, 2008.

\bibitem{wang2021understanding}
F.~Wang and H.~Liu, ``Understanding the behaviour of contrastive loss,'' in
  \emph{Proceedings of the IEEE/CVF conference on computer vision and pattern
  recognition}, 2021, pp. 2495--2504.

\bibitem{chen2024power}
Z.~Chen, Y.~Fang, Y.~Zhang, L.~Guo, J.~Chen, H.~Chen, and W.~Zhang, ``The power
  of noise: Toward a unified multi-modal knowledge graph representation
  framework,'' \emph{arXiv preprint arXiv:2403.06832}, 2024.

\bibitem{zhang2020learning}
Z.~Zhang, J.~Cai, Y.~Zhang, and J.~Wang, ``Learning hierarchy-aware knowledge
  graph embeddings for link prediction,'' in \emph{Proceedings of the AAAI
  conference on artificial intelligence}, vol.~34, no.~03, 2020, pp.
  3065--3072.

\bibitem{li2023knowledge}
L.~Li, X.~Zhang, Z.~Jin, C.~Gao, R.~Zhu, Y.~Liang, and Y.~Ma, ``Knowledge graph
  completion method based on quantum embedding and quaternion interaction
  enhancement,'' \emph{Information Sciences}, vol. 648, p. 119548, 2023.

\bibitem{shang2023relation}
B.~Shang, Y.~Zhao, D.~Wang, and J.~Liu, ``Relation-aware multi-positive
  contrastive knowledge graph completion with embedding dimension scaling,'' in
  \emph{Proceedings of the 46th International ACM SIGIR Conference on Research
  and Development in Information Retrieval}, 2023, pp. 878--888.

\bibitem{zhang2023weighted}
Z.~Zhang, Z.~Guan, F.~Zhang, F.~Zhuang, Z.~An, F.~Wang, and Y.~Xu, ``Weighted
  knowledge graph embedding,'' in \emph{Proceedings of the 46th international
  ACM SIGIR conference on research and development in information retrieval},
  2023, pp. 867--877.

\bibitem{li2024sphere}
Z.~Li, Y.~Ao, and J.~He, ``Sphere: Expressive and interpretable knowledge graph
  embedding for set retrieval,'' in \emph{Proceedings of the 47th International
  ACM SIGIR Conference on Research and Development in Information Retrieval},
  2024, pp. 2629--2634.

\bibitem{li2025multi}
L.~Li, Z.~Jin, X.~Zhang, H.~Duan, J.~Wang, Z.~Tao, H.~Zhao, and X.~Zhu,
  ``Multi-view riemannian manifolds fusion enhancement for knowledge graph
  completion,'' \emph{IEEE Transactions on Knowledge and Data Engineering},
  2025.

\bibitem{liangclustering}
K.~Liang, Y.~Liu, H.~Li, L.~Meng, S.~Liu, S.~Wang, X.~Liu \emph{et~al.},
  ``Clustering then propagation: Select better anchors for knowledge graph
  embedding,'' in \emph{The Thirty-eighth Annual Conference on Neural
  Information Processing Systems}.

\bibitem{pavlovicexpressive}
A.~Pavlovi{\'c} and E.~Sallinger, ``Expressive: A spatio-functional embedding
  for knowledge graph completion,'' in \emph{The Eleventh International
  Conference on Learning Representations}.

\bibitem{chen2023dipping}
C.~Chen, Y.~Wang, A.~Sun, B.~Li, and K.-Y. Lam, ``Dipping plms sauce: Bridging
  structure and text for effective knowledge graph completion via conditional
  soft prompting,'' \emph{arXiv preprint arXiv:2307.01709}, 2023.

\bibitem{pmlr-v202-lee23c}
J.~Lee, C.~Chung, and J.~J. Whang, ``{I}n{G}ram: Inductive knowledge graph
  embedding via relation graphs,'' in \emph{Proceedings of the 40th
  International Conference on Machine Learning}, ser. Proceedings of Machine
  Learning Research, A.~Krause, E.~Brunskill, K.~Cho, B.~Engelhardt, S.~Sabato,
  and J.~Scarlett, Eds., vol. 202.\hskip 1em plus 0.5em minus 0.4em\relax PMLR,
  23--29 Jul 2023, pp. 18\,796--18\,809.

\bibitem{shang2024mixed}
B.~Shang, Y.~Zhao, J.~Liu, and D.~Wang, ``Mixed geometry message and trainable
  convolutional attention network for knowledge graph completion,'' in
  \emph{Proceedings of the AAAI Conference on Artificial Intelligence},
  vol.~38, no.~8, 2024, pp. 8966--8974.

\bibitem{cao2022geometry}
Z.~Cao, Q.~Xu, Z.~Yang, X.~Cao, and Q.~Huang, ``Geometry interaction knowledge
  graph embeddings,'' in \emph{Proceedings of the AAAI Conference on Artificial
  Intelligence}, vol.~36, no.~5, 2022, pp. 5521--5529.

\bibitem{chen2022hybrid}
X.~Chen, N.~Zhang, L.~Li, S.~Deng, C.~Tan, C.~Xu, F.~Huang, L.~Si, and H.~Chen,
  ``Hybrid transformer with multi-level fusion for multimodal knowledge graph
  completion,'' in \emph{Proceedings of the 45th international ACM SIGIR
  conference on research and development in information retrieval}, 2022, pp.
  904--915.

\bibitem{touvron2023llama}
H.~Touvron, T.~Lavril, G.~Izacard, X.~Martinet, M.-A. Lachaux, T.~Lacroix,
  B.~Rozi{\`e}re, N.~Goyal, E.~Hambro, F.~Azhar \emph{et~al.}, ``Llama: Open
  and efficient foundation language models,'' \emph{arXiv preprint
  arXiv:2302.13971}, 2023.

\bibitem{wang2021visual}
M.~Wang, S.~Wang, H.~Yang, Z.~Zhang, X.~Chen, and G.~Qi, ``Is visual context
  really helpful for knowledge graph? a representation learning perspective,''
  in \emph{Proceedings of the 29th ACM International Conference on Multimedia},
  2021, pp. 2735--2743.

\bibitem{liang2023knowledge}
K.~Liang, Y.~Liu, S.~Zhou, W.~Tu, Y.~Wen, X.~Yang, X.~Dong, and X.~Liu,
  ``Knowledge graph contrastive learning based on relation-symmetrical
  structure,'' \emph{IEEE Transactions on Knowledge and Data Engineering},
  vol.~36, no.~1, pp. 226--238, 2023.

\bibitem{liang2024simple}
K.~Liang, L.~Meng, Y.~Liu, M.~Liu, W.~Wei, S.~Liu, W.~Tu, S.~Wang, S.~Zhou, and
  X.~Liu, ``Simple yet effective: Structure guided pre-trained transformer for
  multi-modal knowledge graph reasoning,'' in \emph{Proceedings of the 32nd ACM
  International Conference on Multimedia}, 2024, pp. 1554--1563.

\bibitem{liu2024dysarl}
K.~Liu, F.~Zhao, Y.~Yang, and G.~Xu, ``Dysarl: Dynamic structure-aware
  representation learning for multimodal knowledge graph reasoning,'' in
  \emph{Proceedings of the 32nd ACM International Conference on Multimedia},
  2024, pp. 8247--8256.

\bibitem{mousselly2018multimodal}
H.~Mousselly-Sergieh, T.~Botschen, I.~Gurevych, and S.~Roth, ``A multimodal
  translation-based approach for knowledge graph representation learning,'' in
  \emph{Proceedings of the Seventh Joint Conference on Lexical and
  Computational Semantics}, 2018, pp. 225--234.

\end{thebibliography}
\bibliographystyle{IEEEtran}


\vfill

\end{document}